\documentclass{article}
\usepackage{graphicx}  
\usepackage{amsmath}   
\usepackage[compress]{cite}
\usepackage{amssymb}   
\usepackage{bm} 
\usepackage{dcolumn}
\usepackage{color}
\usepackage{comment}
\usepackage{mathrsfs}
\usepackage{amsfonts}
\usepackage{varioref}
\usepackage{csquotes}
\usepackage{float}
\usepackage[caption=false]{subfig}
\RequirePackage[colorlinks,citecolor=blue,urlcolor=magenta,linkcolor=blue]{hyperref}
\def\KN{Kerr-Newmann }

\def\KN{Kerr-NUT }

\def \GH {Geroch-Hansen}

\labelformat{chapter}{Chapter (#1)} 
\labelformat{section}{Section #1} 
\labelformat{subsection}{Section #1} 
\labelformat{subsubsection}{Section #1}
\labelformat{subsubsubsection}{Section #1}
\labelformat{equation}{Eq.~(#1)} 
\labelformat{figure}{Fig.~#1} 
\labelformat{subfigure}{Fig.~\thefigure#1} 
\labelformat{table}{Tab.~#1} 
\labelformat{appendix}{Appendix #1}
\title{Multipole moments of compact objects with NUT charge: Theoretical and observational implications}
\author{Sajal Mukherjee\footnote{sajal@iucaa.in}~$^{1}$ and Sumanta Chakraborty\footnote{sumantac.physics@gmail.com}~$^{2}$
\\
$^{1}${\small{Inter-University Centre for Astronomy and Astrophysics, Post Bag 4, Pune-411007, India}}
\\
$^{2}${\small{School of Physical Sciences}}
\\
{\small{Indian Association for the Cultivation of Science, Kolkata-700032, India}}
}
\begin{document}

\maketitle
\begin{abstract}
We derive the multipole moments of the Kerr-NUT black hole spacetime using the Geroch-Hansen formalism, even though the spacetime is not asymptotically flat. Intriguingly, in the presence of the NUT charge, the absence of reflection symmetry about the equatorial plane, leads to mass and spin multipole moments of all orders, in stark contrast to Kerr-like spacetimes. This leads to a drastic departure of the multipolar structure of a compact object with NUT charge, whose implications for gravitational wave observations have been explored. Our analysis of multipole moments for the Kerr-NUT spacetime is also in tune with the Thorne's approach.  
\end{abstract}
\section{Introduction}

A gravitating system is most often characterized by its multipole moments. These moments are intimately connected with the intrinsic properties of the gravitating object and hence makes them very useful as well as important in astrophysics. This is because, determination of the multipole moments will yield handful of information about the nature and properties of the gravitating object. In particular, the structure of the multipole moments can be used to distinguish black holes from other compact objects, e.g., neutron stars and will provide an experimental verification of the no-hair theorem \cite{Cardoso:2016ryw,Johannsen:2011dh}. 

Recent discovery of the gravitational waves \cite{Abbott:2020khf,Abbott:2020uma,LIGOScientific:2018mvr,Abbott:2017gyy,TheLIGOScientific:2017qsa,Abbott:2016blz,TheLIGOScientific:2016pea,Chakraborty:2017qve,Chakravarti:2018vlt,Chakravarti:2019aup} has further boosted the research on multipole moments of gravitating objects as they play a crucial role in proper understanding of the gravitational wave observations. This has resulted into a broad spectrum of works relating the theoretical computation of multipole moments with observational aspects \cite{Bianchi:2020bxa}. The theoretical backdrop for these searches regarding multipole moments of gravitating objects are primarily based on the works lead by Geroch \cite{Geroch:1970cc,Geroch:1970cd}, Hansen \cite{Hansen:1974zz} and Thorne \cite{Thorne:1980ru}. These seminal works prepare the ground for most of the recent works on multipole moments of gravitating objects, see e.g., \cite{Pook-Kolb:2020jlr,Vigeland:2010xe}. Recently, the gravitational wave observations has brought these multipole moments to the forefront of gravitational wave astronomy, with direct observational implications in the astrophysical as well as astronomical realms \cite{Datta:2019euh}. Possible implications of these multipole moments in the upcoming Laser Interferometer Space Antenna (LISA) have been proposed in Refs. \cite{Ryan:1997hg,Barack:2006pq}, while study of these moments and their evolution during merger has been studied using numerical relativity in \cite{Cabero:2014nza,Prasad:2020xgr}. On the theoretical side as well, multipole moments of a gravitating object in higher spacetime dimensions have been studied in \cite{Shiromizu:2004jt}, while Refs. \cite{Compere:2017wrj,Pappas:2018csu} discuss the formalism to obtain multipole moments of a gravitating object in alternative theories of gravity. In what follows we will provide a brief introduction to the Geroch-Hansen and Thorne multipole moments, which will be useful for our later purposes.  

In the Newtonian theory, it is straightforward to define the multipole moments of a mass distribution by expanding its gravitational potential in an asymptotic series in the inverse power of the radial distance from the gravitating object. Various multipole moments of the mass distribution can be read off from various powers of the radial distance of the field point from the mass distribution. In this Newtonian description, the key ingredient is the condition of asymptotic flatness, i.e., the potential must vanish at a field point, which is located at a large distance from the source. An extension of the Newtonian description of multipole moments to general relativity was a major concern related to the lack of covariance in the definition of multipole moments as well as a suitable implementation of the asymptotic flatness. The notion of asymptotic flatness can be imposed in a simple manner, by ensuring that at spatial infinity, the metric $g_{\mu \nu}$ reduces to flat spacetime metric $\eta_{\mu \nu}$. However, the multipole moments can be defined in a covariant manner in stationary spacetimes, which inherit a timelike Killing vector field. The covariant approach was first described by Geroch for static spacetimes \cite{Geroch:1970cd} and it was used to derive the associated multipole moments. It was subsequently extended by Hansen \cite{Hansen:1974zz} for stationary spacetimes and thus collectively it is referred to as the \GH\ formalism. As the previous discussion suggests, the \GH\ formalism applies to stationary spacetimes alone. In generic circumstances, e.g., during the in-spiral of two compact objects, the stationarity assumption breaks down. In this context, another approach to obtain the multipole moments was proposed by Thorne in Ref. \cite{Thorne:1980ru}, where one introduces a coordinate system known as asymptotically Cartesian and mass centered (ACMC) and then expands the metric elements in a power series in the distance of the field point from the origin of the ACMC coordinate system. Interestingly, as shown in Ref. \cite{gursel1983multipole}, both the \GH\ formalism and the Thorne formalism are equivalent upto some normalization. While both these formalisms predict identical results, there exist qualitative differences between them. In particular, the moments computed by the \GH\ formalism arises from a generally covariant prescription, which is not the case for the Thorne's approach. On the other hand, the \GH\ formalism applies to stationary spacetimes alone, but Thorne's approach is much more general than that. For a general review on the multipole moments, we refer the reader to \cite{quevedo1990multipole}. 

In the present work, we attempt to obtain the multipole moments of the \KN spacetime, which is a vacuum solution of the Einstein's field equations and describes a stationary, axi-symmetric and asymptotically non-flat spacetime \cite{Newman:1963yy}, in the sense that the metric for the \KN spacetime cannot be reduced to flat spacetime metric, asymptotically. There are two major motivations to study this problem --- first of all we would like to understand whether the non-trivial asymptotic limit of the \KN spacetime is a hindrance towards application of the \GH\ multipole moments, given that \KN spacetime is stationary. Secondly, we would like to explore the observational implications of the NUT charge and the gravitational wave observations seems to be the best avenue to probe the same. We will use some intriguing structure, the multipole moments of a compact object would inherit in the presence of a NUT charge, which we will explicitly demonstrate using the \KN spacetime. So far, there are several attempts to look for an avenue, where the very existence of the NUT charge can be determined \cite{Chakraborty:2019rna,Chakraborty:2017nfu}, see also \cite{Manko:2005nm,Manko:2006bf,Jefremov:2016dpi}, and \cite{Clement:2020gjy,Bordo:2019rhu,Mukherjee:2018dmm,Rahman:2020guv} for some recent works. But none of these studies, except \cite{Rahman:2020guv} involved gravitational wave observations and thus the results presented in this work will provide a new insight into the problem. In the present work, we aim to revisit the formalism developed in \cite{Ryan:1995wh,fodor1989multipole}, where several gravitational wave observables were directly related to the multipole moments of the central compact object during an inspiral phase, but in the context of the \KN spacetime. As we will see, the presence of the NUT charge modifies the formalism significantly and thus provides a very direct test to probe the NUT charge using gravitational wave observations.  

The manuscript is organized as follows: We start with a brief introduction about the \KN spacetime in \ref{Sec_KN}, where we have expressed the \KN metric in various coordinate systems and have discussed the asymptotic behaviour of the same. Following this discussion, we have worked out in detail the individual steps for the determination of the multipole moments and have derived the moments of the \KN black hole spacetime in \ref{Sec_Mult_Mom_KN}. Application of these moments to gravitational wave astronomy has been explored in \ref{KN_GW}. Finally, we conclude with a discussion on our results. In addition, we have also completed the discussion in the main text with certain computations presented in the appendices. 

\textit{Notations and Conventions:} Throughout this paper, we use the geometrical units, $c=1=G$ for convenience. Greek indices, $\mu,\nu,\ldots$ are used to denote four dimensional spacetime indices, while Roman indices $a,b,\ldots$ are used to denote indices on the induced lower dimensional hypersurface. Further, we use mostly positive signature convention, i.e., the flat spacetime metric is taken to be $\eta_{\mu \nu}=\textrm{diag}(-1,1,1,1)$.  

\section{Structure of the \KN spacetime}\label{Sec_KN}

In this section, we will discuss the geometrical structure of the \KN spacetime, which will be central to our discussion of the multipole moments. The \KN spacetime can be expressed in several different coordinate systems, one of which is of course the standard Boyer-Lindquist coordinate system. In which, the line element for the \KN spacetime can be expressed as \cite{Newman:1963yy,LyndenBell:1996xj,Chen:2006xh,Awad:2005ff,Dehghani:2005zm,Mukhopadhyay:2003iz,Dadhich:2001sz},
\begin{align}\label{metric_KN_old}
ds^{2}=-\frac{\Delta}{\Sigma^{2}}\left(dt-Pd\phi\right)^{2}
+\frac{\sin ^{2}\theta}{\Sigma^{2}}\left\{\left(r^{2}+a^{2}+N^{2}\right)d\phi -adt\right\}^{2}
+\frac{\Sigma^{2}}{\Delta}dr^{2}+\Sigma^{2}d\theta ^{2}~,
\end{align}
where the quantities $\Delta$, $P$ and $\Sigma^{2}$ have the following expressions,
\begin{align}\label{def_eq}
\Delta \equiv r^{2}-2Mr+a^{2}-N^{2}~;\qquad P \equiv a\sin ^{2}\theta -2N\cos \theta~;\qquad \Sigma^{2} \equiv r^{2}+(N+a\cos \theta)^{2}~.
\end{align}
Here, $M$ stands for the mass of the black hole, $a$ is the rotation parameter and $N$ corresponds to the NUT charge. In the limit of vanishing NUT charge, the above metric reduces to the Kerr spacetime and in the limit of vanishing NUT charge and rotation parameter we get back the Schwarzschild spacetime. As we will see in the later sections, this metric will be central in our determination of the multipole moments for the Kerr-NUT spacetime. 

We must emphasize that the Kerr-NUT spacetime in Boyer-Lindquist coordinate system inherits an interesting duality property \cite{Turakulov:2001jc}. The Kerr metric, obtained by substituting $N=0$ in the \KN spacetime, is dual to the metric obtained by substituting $M=0$ in \ref{metric_KN_old}, where the duality implies the following transformation, $M\leftrightarrow iN$ and $r\leftrightarrow ia\cos\theta$. As we will see, even the multipole moment analysis will respect this symmetry, thus depicting another interesting aspect of the Kerr-NUT spacetime. Note that the duality symmetry requires a non-zero rotation parameter \cite{Argurio:2009xr}.

In describing the Kerr-NUT spacetime, it is also worthwhile to touch upon another useful coordinate system, namely the cylindrical polar coordinate system. In contrast to the Boyer-Lindquist coordinate system $(t,r,\theta,\phi)$ depicted above, the cylindrical polar coordinate system can be described using the $(t,\rho,z,\phi)$ coordinates, where $t$ and $\phi$ are identical to the respective coordinates in the Boyer-Lindquist system. The other two coordinates $\rho$ and $z$ are related to the Boyer-Lindquist coordinates $r$ and $\theta$, through the following relation \cite{kramer1987ernst}, 
\begin{align}\label{KN_transform}
\rho=\sqrt{\Delta(r)}\sin \theta~;\quad z=\left(r-M\right)\cos\theta~,
\end{align}
where $\Delta(r)$ has been defined in \ref{def_eq}. For a derivation of the above coordinate transformation, see \ref{AppSphCy}. Using the above coordinate transformations between the Boyer-Lindquist coordinates $(r,\theta)$ and the cylindrical coordinates $(\rho,z)$, we can write down the \KN metric in the cylindrical polar coordinate system, in which the line element becomes,
\begin{align}\label{KN_Cylindrical}
ds^{2}=-F(\rho,z)\Big[dt-\bar{\omega}(\rho,z)d\phi\Big]^{2}+\frac{1}{F(\rho,z)}\left[e^{2\gamma(\rho,z)}\left(d\rho^{2}+dz^{2}\right)+\rho^{2}d\phi^{2}\right]~.
\end{align}
The above form of the \KN line element involves three functions, namely $F(\rho,z)$, $\bar{\omega}(\rho,z)$ and $\gamma(\rho,z)$, respectively. Each of these functions can be expressed in terms of the quantities defined in \ref{def_eq} as,
\begin{align}
F=\frac{\Delta-a^{2}\sin^{2}\theta}{\Sigma^{2}}~;\quad \bar{\omega}=\frac{\Delta P-a\left(r^{2}+a^{2}+N^{2}\right)\sin^{2}\theta}{\Delta-a^{2}\sin^{2}\theta}~;
\quad e^{2\gamma}=\frac{\Sigma^{2}}{\Delta \cos^{2}\theta+(r-M)^{2}\sin^{2}\theta}~.
\end{align}
For the determination of the energy loss through gravitational radiation and also for finding out the characteristic frequencies associated with the motion of the compact object on the equatorial plane, during its inspiral around a Kerr black hole, the above cylindrical coordinate system is often employed. In addition, note that in the asymptotic limit (i.e., $r\rightarrow \infty$), these coordinates behave as normal cylindrical coordinates, $\rho\rightarrow r\sin \theta$ and $z\rightarrow r\cos \theta$, respectively. This asymptotic structure of the cylindrical coordinates will have implications in the subsequent derivation of the multipole moment for the Kerr-NUT black hole. 

The singularity structure of the Kerr-NUT spacetime also needs an adequate discussion. There is no singularity at $r=0$, but there is a singularity on the axis of symmetry, often referred to as the Misner string \cite{Bordo:2019tyh}. There are two possible interpretations for this singular structure, the first one is due to Misner \cite{misner1963flatter} (see also \cite{Clement:2015cxa,Clement:2015aka}), where the singularity can be avoided but at the cost of the introduction of a periodic time coordinate. As a consequence, there exist closed timelike curves through every event in the spacetime and questions the causality structure of the spacetime. The other interpretation, due to Bonnor \cite{bonnor1969new} is more appealing from the physical point of view, where the singularity was attributed to a massless rotating rod. This is the interpretation we provide to the metric considered here as well. It is certainly possible to re-distribute the singularity over the axis by introducing certain additional parameters in the problem, as advocated in \cite{Manko:2005nm}. But it will not affect the results presented in this work, since our motivation was to study multipole moments of this metric, which involves behaviour of the spacetime at a large distance, where the nature of singularity does not play any significant role. This is further supported by several recent works as well, where various astrophysical properties of the NUT solution have been studied in detail \cite{Jefremov:2016dpi,Cebeci:2015fie,Pradhan:2014zia,Mukhopadhyay:2003iz,Mukherjee:2018dmm}.

Let us now discuss another interesting and equally intriguing feature of this metric, namely the above line element is asymptotically \emph{non-flat}, since it cannot be reduced to the flat spacetime metric $\eta_{\mu \nu}$ using any coordinate transformation. This is clear from the asymptotic ($r\rightarrow \infty$) expression of \ref{metric_KN_old}, which takes the following form,
\begin{equation}\label{asymp_KN}
ds^{2}=-dt^{2} \left[1+4N\cos\theta \dfrac{d\phi}{dt}\right]+dr^{2}+r^2 (d\theta^2+\sin^2\theta d\phi^2)~.
\end{equation}
Presence of the $dtd\phi$ cross term gives rise to a rotation-like effect at infinity and is solely dependent on the NUT charge. Explicit computation of the Riemann tensor components from \ref{metric_KN_old} shows that asymptotically all of them vanishes as $\mathcal{O}(1/r^{3})$ as spatial infinity is approached, but redefinition of $t$ or $\phi$ cannot reduce \ref{asymp_KN} to a flat form for the metric. This is why we cannot consider the NUT charge as originating from a localized source embedded in an originally flat spacetime \cite{Misner:1963fr}. The reason behind the existence of the $dtd\phi$ cross term in the asymptotic form for the Kerr-NUT geometry is deeply rooted in the gravitomagnetic origin of the NUT charge, while other parameters, such as the mass or electric charge gives rise to gravitoelectrical effects \cite{LyndenBell:1996xj}. This also connects up well with the duality symmetry between the gravitoelectric and the gravitomagnetic parts, in which the NUT charge is dual to the mass parameter along with an exchange between radial and angular coordinates \cite{Turakulov:2001jc}. Since the multipole structure of a spacetime depends crucially on the nature of the metric at asymptotic infinity, it remains to see whether the available formalisms for determination of the multipole moment is applicable in the present context. This is what we explore in the next section. 

\section{Multipole moments of a Kerr-NUT black hole}\label{Sec_Mult_Mom_KN}

Having briefly reviewed the key ingredients and properties of a Kerr-NUT black hole, in this section we would like to derive the multipole moments of a Kerr-NUT black hole, where, we revisit the mechanism to obtain the multipole moments which are otherwise known for Kerr
spacetime. As we will see, for Kerr-NUT black hole the multipole moments can be expressed in a very compact form, akin to the case of Kerr black hole, with several interesting properties. These will be useful while discussing the implications of a NUT charge for gravitational wave astronomy. As in the case of a Kerr black hole, here also the key steps to compute the multipole moments are as follows: (a) deriving the twist potential for the \KN spacetime, (b) projecting to a lower dimensional hypersurface, which is asymptotically flat, (c) hence obtaining the Ernst potential leading to the multipole moments. For clarity we will present all these steps explicitly in the ensuing discussion. 
\subsection{Deriving the twist potential} 

Since the \KN black hole depicts a stationary spacetime, there is a timelike Killing vector field $\xi^{\mu}_{\rm (t)}\equiv (\partial/\partial t)^{a}$ associated with it. Using this time-like Killing vector field we can define two quantities, namely the norm $\lambda$ and twist potential $\omega$, which will be of significant use lately in this paper. For the \KN black hole, the norm $\lambda$ of the time-like Killing vector field $\xi^{\mu}_{\rm (t)}$ is given by, 
\begin{equation}\label{KN_norm}
\lambda \equiv \xi^{\mu}_{\rm (t)}\xi_{\mu}^{\rm (t)}=-g_{tt}=\dfrac{1}{\rho^2} \left[\Delta(r)-a^2 \sin^2\theta\right]~,
\end{equation}
where, the function $\Delta(r)$ has been defined in \ref{def_eq} and depends on all the ``hairs" of the black hole, namely the mass $M$, the rotation parameter $a$ and the NUT charge $N$, respectively. We will have several occasions to use this norm in the subsequent sections. On the other hand, the twist vector field $\omega_{\mu}$ is defined in terms of the Killing vector field $\xi^{\mu}_{\rm (t)}$ as follows,
\begin{equation}\label{KN_twistvec}
\omega_{\mu}=\sqrt{-g}\epsilon_{\mu \nu \rho \sigma}\xi_{\rm (t)}^{\nu}\nabla^{\rho}\xi_{\rm (t)}^{\sigma}~,
\end{equation}
where, $\nabla$ is the standard covariant derivative operator, $\xi^{\mu}_{\rm (t)}$ is the time-like killing vector field of the \KN~spacetime defined earlier and $\epsilon_{\mu \nu \rho \sigma}$ is the completely antisymmetric Levi-Civita symbol, with $\epsilon_{0123}=+1$. The twist potential $\omega$ arises out of the twist vector defined above, such that, $\omega_{\mu}\equiv \nabla_{\mu}\omega$. It is worthwhile to emphasize that even though the vector $\omega_{\mu}$ can be defined in any spacetime having time-like Killing vector field, the twist potential $\omega$ may not exist. Since the existence of $\omega$ requires $\omega_{\mu}$ to be hypersurface orthogonal, which is not generically true. It follows that for vacuum spacetimes, $\omega_{\mu}$ is indeed hypersurface orthogonal and hence the twist potential is guaranteed to exist (for a derivation, see \ref{AppTwist}). Since the Kerr-NUT black hole is also a vacuum solution of Einstein gravity, the twist potential $\omega$ will exist and can be determined by an explicit computation, presented below. 

As an aside, let us briefly discuss why we are using the Killing vector field $\xi^{\mu}_{\rm (t)}=(\partial/\partial t)^{\mu}$ rather than $\xi^{\mu}_{\rm H}=(\partial/\partial t)^{\mu}+\Omega_{\rm H}(\partial/\partial \phi)^{\mu}$, even though both are time-like in the asymptotic region. This is intimately connected with the fact that $\xi_{\rm (t)}^{\mu}$ is not hypersurface orthogonal, while $\xi^{\mu}_{\rm H}$ is. As a consequence, by Frobenius theorem \cite{wald2010general} it follows that, $\xi^{[\alpha}_{\rm H}\nabla^{\beta}\xi^{\gamma]}_{\rm H}=0$ and hence the twist vector field associated with the Killing vector $\xi^{\mu}_{\rm H}$ identically vanishes. Thus in order to have a non-zero twist vector field and hence a non-zero twist potential, it is necessary to work with the Killing vector field $\xi^{\mu}_{\rm (t)}$ instead. In brief, $\omega_{\alpha}$ measures the failure of the Killing vector field $\xi^{\mu}_{\rm (t)}$ to become hypersurface orthogonal. 

Returning back to the computation of the twist potential in the \KN black hole spacetime, we start by explicitly evaluating the twist vector field $\omega_{\alpha}$, defined in \ref{KN_twistvec}. Among the four components, $\omega_{t}$ and $\omega_{\phi}$ identically vanishes. The vanishing of $\omega_{t}$ is obvious from \ref{KN_twistvec}, as it immediately follows from the antisymmetry of the Levi-Civita symbol that $\omega_{\alpha}\xi_{\rm (t)}^{\alpha}=0$. Furthermore, we also have, $\omega_{\alpha}(\partial/\partial \phi)^{\alpha}=0$, which follows from the result that $\xi^{\mu}_{\rm (t)}$ and $\xi^{\mu}_{(\phi)}\equiv (\partial/\partial \phi)^{a}$ are commuting Killing vector fields \cite{wald2010general}.

Among the other non-vanishing components of the twist vector, the explicit expression for $\omega_{\theta}$, in terms of the metric components and their derivatives, takes the following form,
\begin{equation}\label{omega_theta}
\omega_{\theta}=\partial_{\theta}\omega=\sqrt{-g}g^{rr}\left(g^{\phi \phi}\partial_{r}g_{t\phi}+g^{t\phi}\partial_{r}g_{tt}\right)~.
\end{equation}
In the present context of the Kerr-NUT black hole spacetime, the expressions for $\partial_{r}g_{t\phi}$ and $\partial_{r}g_{tt}$ can be computed in a straightforward manner from \ref{metric_KN_old} and takes the following form,
\begin{eqnarray}
\partial_{r}g_{t\phi}&=&\Sigma^{-2}\left[(2 r-2 M)P-2 ar \sin^2\theta\right]-2r\Sigma^{-4}\left[\Delta P-a \sin^2\theta(r^2+a^2+N^2)\right]~,  
\nonumber 
\\
\partial_{r}g_{tt}&=&-\Sigma^2 (2r-2M)+2r\Sigma^{-4}\left(\Delta-a^2 \sin^2\theta\right)~.
\end{eqnarray}
Substituting these expressions for $\partial_{r}g_{t\phi}$ and $\partial_{r}g_{tt}$, along with the rest of the metric components for the Kerr-NUT metric, in \ref{omega_theta}, we obtain,
\begin{align}
\omega_{\theta}&=\Sigma^{-4}(\sin\theta)^{-1}\Big\{(\Delta-a^2 \sin^2\theta)\left[(2r-2M)P-2ar \sin^2\theta\right]
\nonumber 
\\
&\hskip 3 cm -(2r-2M)\bigl[\Delta P-a\sin^2\theta(r^2+a^2+N^2)\bigr]\Big\}
\nonumber
\\
&=\dfrac{2Ma \sin\theta\left(r^2+N^2-a^2 \cos^2\theta\right)}{\Sigma^4}+\dfrac{4aN\sin\theta}{\Sigma^4}(r-M)\left(N+a\cos\theta\right)~.
\label{eq:omega_theta}
\end{align}
Along identical lines, the expression for $\omega_{r}$ can also be computed in terms of the metric and its derivatives, yielding
\begin{equation}
\omega_{r}=\partial_{r}\omega=-\sqrt{-g}g^{\theta \theta}\left\{g^{\phi \phi}\partial_{\theta}g_{t\phi}+g^{t\phi}\partial_{\theta}g_{tt}\right\}.
\end{equation}
For the Kerr-NUT black hole spacetime, the expressions for $\partial_{\theta}g_{t\phi}$ and $\partial_{\theta}g_{tt}$ takes the following form,
\begin{align}
\partial_{\theta} g_{t\phi}&=\frac{\sin\theta}{\Sigma^{2}} \Bigl[\Delta (2 a \cos\theta+2N)-2a\cos\theta(r^2+a^2+N^2) 
\nonumber 
\\
&\hskip 2 cm +\Sigma^{-2}\Bigl\{2a\left(N+a \cos\theta\right)\left[\Delta P-a \sin^2\theta \left(r^2+a^2+N^2\right)\right]\Bigr\}\Bigr]~, 
\nonumber 
\\
\partial_{\theta} g_{tt}&= \dfrac{\sin\theta}{\Sigma^2} \Bigl[2a^2 \cos\theta-\frac{a}{\Sigma^{2}}\left(N+a \cos\theta\right)\left(\Delta-a^2 \sin^2\theta\right)\Bigr]~.
\end{align}
Again, substituting the respective expressions for $\partial_{\theta}g_{t\phi}$, $\partial_{\theta}g_{tt}$ and other metric elements, the expression for $\omega_{r}$ in the \KN black hole spacetime takes the following form,
\begin{equation}
\omega_{r}=\partial_{r}\omega=\dfrac{1}{\Sigma^4}\Bigl[4Mar\cos\theta-2N(r^2-2Mr)+2N(N+a\cos\theta)^2\Bigr]~.
\label{eq:omega_r}
\end{equation}
As emphasized earlier, the \KN spacetime being a vacuum solution of the Einstein's equations, guarantees the existence of a twist potential. Also the solution being stationary and axi-symmetric, demands $\omega=\omega(r,\theta)$. Thus \ref{eq:omega_theta} and \ref{eq:omega_r} provides the two partial differential equations necessary to solve for $\omega$. Solving these equations we obtain the following expression for the twist potential $\omega$ in the \KN spacetime,
\begin{equation}
\omega=-\dfrac{2Ma \cos\theta}{\Sigma^2}+\dfrac{2N(r-M)}{\Sigma^2}~.
\label{eq:twist_pot}
\end{equation}
Note that in the limit of vanishing NUT charge (i.e., $N=0$) the twist potential presented above reduces to the respective expression for the Kerr black hole \cite{Vigeland:2010xe}, as desired. Interestingly, even in the limit of vanishing rotation parameter (i.e., $a=0$) the twist potential is non-zero and is proportional to the NUT charge. This expression corresponds to the twist potential for the Schwarzschild-NUT spacetime, another vacuum solution to general relativity. Interestingly, the twist potential with zero NUT charge is related to the twist potential with zero mass through the duality transformation, $M\leftrightarrow iN$ and $r\leftrightarrow ia\cos\theta$, as one can immediately check from \ref{eq:twist_pot}. Thus the twist potential obeys the duality transformation property of the \KN spacetime. This finishes the first part of the story, as we have derived the norm of the time-like Killing vector field and the twist potential associated with it. It is now time to introduce the metric on the lower dimensional manifold and hence derive the conformal completion of the spatial sector. 
\subsection{Projecting to a lower dimensional manifold}\label{lower_dim_KN}

As emphasized earlier, the four-dimensional \KN geometry is not asymptotically flat due to the non-vanishing contribution of the $g_{t\phi}$ component from the NUT charge at spatial infinity. Thus at first sight it may seem that the Geroch-Hansen formalism will not be directly applicable. However, as we will show, there exists a three dimensional manifold $M_{3}$, which is asymptotically flat and is sufficient to define the multipole structure of the spacetime through the Geroch-Hansen formalism. In the following segments, we will discuss this aspect in detail. 

Let us start by introducing the projector $h_{\mu \nu}$, such that $h_{\mu \nu}\xi^{\mu}_{\rm (t)}=0$, which can be expressed in terms of the background metric $g^{\mu \nu}$ as,
\begin{equation}
h_{\mu \nu}=\lambda g_{\mu \nu}+\xi^{\rm (t)}_{\mu}\xi_{\nu}^{\rm (t)}~.
\end{equation}
One can easily verify that the components $h_{tt}$ and $h_{t\phi}$ of the projector $h_{\mu \nu}$ identically vanishes, while the spatial components are non zero and take the following forms
\begin{equation}
h_{rr}=\dfrac{1}{\Delta}\left(\Delta-a^2 \sin^2\theta\right)~; \quad h_{\theta \theta}=\left(\Delta-a^2 \sin^2\theta\right)~; \quad h_{\phi \phi}=\Delta \sin^2\theta~.
\end{equation}
This suggests to define a three dimensional manifold $M_{3}$ with coordinates $y^{i}=\{r,\theta,\phi\}$ and metric $h_{ij}=h_{\mu \nu}e^{\mu}_{i}e^{\nu}_{j}$, where $e^{\mu}_{i}=(\partial x^{\mu}/\partial y^{i})$. Since the coordinates of the three dimensional manifold are same as the spatial coordinates of the full spacetime, the metric $h_{ij}$ of the three dimensional manifold is given by, $h_{ij}=\textrm{diag}(h_{rr},h_{\theta \theta},h_{\phi\phi})$. 

For applicability of the Geroch-Hansen formalism, we need the metric $h_{ij}$ to be asymptotically flat. This corresponds to taking the limit $r\rightarrow \infty$, which indeed reduces $h_{ij}$ to $\eta_{ij}$, expressed in spherical polar coordinates. Thus the manifold $M_{3}$ is asymptotically flat. Having identified the asymptotic point $P_{\rm A}$, we define a new manifold $\overline{M_{3}}\equiv M_{3} \cup P_{\rm A}$. This should also define a new metric $\overline{h}_{ab}$ on the three-manifold $\overline{M_{3}}$, such that, $\overline{h}_{ab}=\Omega^{2}h_{ab}$. The conformal factor $\Omega$ must satisfy the three requirements, $\Omega=0$, $\overline{D}_{i}\Omega=0$ and $\overline{D}_{i}\overline{D}_{j}\Omega=2\overline{h}_{ij}$ at the asymptotic infinity, i.e., at the point $P_{\rm A}$. To summarize, we have to re-express the metric on the three-manifold in terms of new coordinates, such that it becomes conformally equivalent to another metric $\overline{h}_{ab}$, with the conformal factor satisfying the properties mentioned above. 

To see how this can be achieved in the present context of Kerr-NUT black hole spacetime, it will turn out to be advantageous to introduce a new radial coordinate $\bar{R}$, which is defined in terms of the old radial coordinate $r$ through the following differential equation,
\begin{equation}\label{rad_eq}
\dfrac{d\bar{R}}{\bar{R}}=-\dfrac{dr}{\sqrt{\Delta(r)}}~,
\end{equation}
where $\Delta(r)$ has been defined in \ref{def_eq}. Note that at large $r$, $\Delta(r)\sim r^{2}$ and hence simple integration of the above differential equation will yield, $\bar{R}\sim (1/r)$. Thus the asymptotic point $P_{\rm A}$, located at $r=\infty$, will map to $\bar{R}=0$ in the new coordinate system. Furthermore, direct integration of the above differential equation, for the \KN black hole spacetime, yields the following relation between the old radial coordinate $r$ and the new radial coordinate $\bar{R}$, 
\begin{equation}\label{coord_transform}
r=M+\frac{1}{\bar{R}}+\dfrac{\bar{R}}{4}\left(M^2-a^2+N^2\right)~.
\end{equation}
As noted earlier, in the limit of $\bar{R}\rightarrow 0$, we obtain $r\rightarrow \infty$. The transformation from the $(r,\theta,\phi)$ to $(\bar{R},\theta,\phi)$ coordinate system yields the following expression for the conformally equivalent metric $\bar{h}_{ab}$,
\begin{equation}\label{con_eq_me}
\bar{h}_{ab}=\Omega^{2}h_{ab}^{2}=\textrm{diag.}\left(1,\bar{R}^2,\bar{R}^2\sin^2\theta e^{-2\beta}\right)~,
\end{equation}
where the conformal factor $\Omega$ takes the following form,
\begin{equation}\label{con_fac}
\Omega=\bar{R}^2\left\{\left[1-\dfrac{1}{4}(M^2-a^2+N^{2})\bar{R}^2\right]^2-a^2 \bar{R}^2 \sin^2\theta\right\}^{-1/2}~.
\end{equation}
In addition, the term $\exp(2\beta)$ appearing in the conformally equivalent metric $\overline{h}_{ab}$, presented in \ref{con_eq_me}, has the following expression in the new coordinate system,
\begin{equation}
e^{2\beta}=1-a^2 \bar{R}^2 \sin^2\theta \left[1-\dfrac{\bar{R}^2}{4} \left(M^{2}-a^{2}+N^{2}\right)\right]^{-2}.
\end{equation}
For our later computations it will be advantageous to express $\beta$ itself in terms of the various parameters of the \KN black hole spacetime in the $(\bar{R},\theta,\phi)$ coordinate system, which after appropriate manipulations yield,
\begin{equation}\label{beta_def}
\beta=\dfrac{1}{2}\ln\left[\dfrac{\bar{\Delta}-a^2 \sin^2\theta}{\bar{\Delta}} \right]~; 
\qquad
\bar{\Delta}=\dfrac{1}{\bar{R}^2}\left[1-\dfrac{\bar{R}^2}{4}\left(M^{2}-a^{2}+N^{2}\right)\right]^{2}.
\end{equation}
Given the conformal factor in \ref{con_fac}, it immediately follows that both $\Omega$ and $\bar{D}_{i}\Omega$ identically vanishes in the $\bar{R}\rightarrow 0$ limit. On the other hand, $\bar{D}_{\bar{R}}\bar{D}_{\bar{R}}\Omega$ is the only non-zero element of $\bar{D}_{i}\bar{D}_{j}\Omega$ at the asymptotic point $P_{\rm A}$, which is consistent with the $\bar{R}\rightarrow 0$ limit of the metric $\overline{h}_{ab}$ on the 3-manifold. Thus we have derived the conformal factor and the conformally equivalent metric on the 3-manifold which will be used in the next section to derive the multipole moments. 

Finally, it is necessary to consider the uniqueness of the conformal factor. Since determination of the multipole moments is intimately connected with the determination of the conformal factor, any arbitrariness in the conformal factor will most likely affect the multipole moments as well. There is indeed some freedom left in the choice of the conformal factor, as we will demonstrate below. Suppose, we make a further conformal transformation, such that the conformal factor scales as, $\Omega\rightarrow \widetilde{\Omega}=e^{\kappa}\Omega$, with $\kappa(P_{\rm A})=0$. This immediately implies, $\widetilde{\Omega}(P_{A})=\Omega(P_{A})$. Using the properties that $\Omega$ satisfy at the asymptotic infinity, it follows that, $\widetilde{\Omega}(P_{\rm A})=0$, as well as, $\bar{D}_{i}\widetilde{\Omega}=e^{\kappa}\bar{D}_{i}\Omega+\widetilde{\Omega}\bar{D}_{i}\kappa$ will vanish at $P_{\rm A}$. In addition, one can also demonstrate that at the asymptotic infinity, i.e., at the point $P_{\rm A}$, $\bar{D}_{i}\bar{D}_{j}\widetilde{\Omega}=2e^{2\kappa}\bar{h}_{ij}$. Therefore, the rescaled conformal factor $\widetilde{\Omega}$ is also a valid candidate to describe the asymptotic structure at infinity. This arbitrariness in the determination of the conformal factor also reflected in the evaluation of the multipole moments and lies in the choice of the origin of the coordinate system. This is an arbitrariness all multipole analysis are plagued with. Often this can be fixed by setting $\nabla_{a}\kappa$ to some pre-assigned value at asymptotic infinity, then it follows that the multipole moments are also fixed \cite{Backdahl:2005be}. This is the route we will also take while determining the multipole moments of the Kerr-NUT black hole spacetime in the next section.

\subsection{The multipole moments from recursion relation}

In this section, we will derive the exact expressions for the multipole moments associated with the Kerr-NUT black hole spacetime using the results derived in the earlier sections. Given the norm $\lambda$ and twist potential $\omega$ associated with the timelike geodesic $\xi^{a}_{\rm (t)}$, one introduces the scalar potential $\Phi$ on the physical manifold, such that,
\begin{equation}\label{potential_eq}
\Phi=\Phi_{\rm M}+i\Phi_{\rm J}~;\quad \Phi_{\rm M}=-\frac{\lambda^{2}+\omega^{2}-1}{(1+\lambda)^{2}+\omega^{2}}~;\quad \Phi_{\rm J}=-\frac{2\omega}{(1+\lambda)^{2}+\omega^{2}}~,
\end{equation}
where, $\Phi_{\rm M}$ is the source for the mass multipole moments and $\Phi_{\rm J}$ is the source of the current multipole moments. For the Kerr-NUT black hole spacetime, expressions for $\lambda$ and $\omega$ can be found from \ref{KN_norm} and \ref{eq:twist_pot} respectively. The above structure of the potential $\Phi$ can actually be derived starting from the complex Ernst potential $\varepsilon$, which in terms of the norm $\lambda$ and the twist potential $\omega$, takes the following form
\begin{equation}\label{ernst_pot}
\varepsilon \equiv \lambda+i\omega~,
\end{equation}
Arising out of which is the potential $\Phi$, expressed in terms of the Ernst potential $\varepsilon$ as,
\begin{equation}\label{mult_pot}
\Phi\equiv \dfrac{1-\varepsilon}{1+\varepsilon}~,
\end{equation}
whose real and imaginary parts are the potentials $\Phi_{\rm M}$ and $\Phi_{\rm J}$ respectively. Given this potential $\Phi$, which is the analog of the Newtonian potential in the non-relativistic context, one will be able to derive the multipole moments by taking successive derivatives of $\Phi$. Since these moments are defined at the asymptotic point $P_{\rm A}$, it is instructive to work with the unphysical metric $\bar{h}_{ab}$ and the unphysical scalar potential $\bar{\Phi}$, which is defined as, $\bar{\Phi}=\Phi/\sqrt{\Omega}$. Thus the Ernst potential in terms of the unphysical scalar potential $\bar{\Phi}$ takes the following form,
\begin{equation}\label{ernst_pot_mod}
\varepsilon=\frac{1-\Phi}{1+\Phi}=\frac{1-\sqrt{\Omega}\bar{\Phi}}{1+\sqrt{\Omega}\Phi}=\frac{\frac{1}{\sqrt{\Omega}}-\bar{\Phi}}{\frac{1}{\sqrt{\Omega}}+\bar{\Phi}}~,
\end{equation}
which will be useful for our later discussion connecting multipole moments with gravitational waves. 

Given the unphysical potential $\bar{\Phi}$, one can read off various multipole moments as derivatives of $\bar{\Phi}$. The monopole term $P$ is a scalar and is simply the potential $\bar{\Phi}$ at the asymptotic point $P_{A}$. Rest of the moments are obtained by taking derivative of the lower moments. In general, the multipole moments are derived from the tensorial recursion relation, which takes the following form,
\begin{align}\label{tensor_recursion}
P&=\bar{\Phi}\vert_{P_{A}}~;\quad P_{a_{1}}=\bar{D}_{a_{1}}\Phi\vert_{P_{A}}~;\quad P_{a_{1}a_{2}}=\textrm{STF}\left[\bar{D}_{a_{1}}P_{a_{2}}-\frac{1}{2}\bar{R}_{a_{1}a_{2}}\right]~;
\nonumber
\\
P_{a_{1}a_{2}\cdots a_{n}}&=\textrm{STF}\left[\bar{D}_{a_{1}}P_{a_{2}\cdots a_{n}}-\frac{(n-1)(2n-3)}{2}\bar{R}_{a_{1}a_{2}}P_{a_{3}\cdots a_{n}} \right]~.
\end{align}
Here, `STF' is the short-form for `Symmetric Trace Free' combination and $\bar{R}_{ab}$ is the Ricci tensor for the unphysical metric $\bar{h}_{ab}$. The above tensorial recursion relation provides the multipole structure of a generic gravitational system. However, for a stationary and axi-symmetric configuration, the moments can also be obtained from a scalar recursion relation \cite{Backdahl:2005be}. In the present case it can be obtained from an even simpler setting, i.e., by repeated differentiation of a scalar function. We will describe the construction of this scalar function for the \KN black hole below. 

Let us start by introducing two new coordinates $\bar{z}$ and $\bar{\rho}$ in terms of the $(\bar{R},\theta)$ coordinates of the unphysical metric $\bar{h}_{ab}$, such that, 
\begin{align}\label{bar_coord}
\bar{z}=\bar{R}\cos \theta~,\quad \bar{\rho}&=\bar{R}\sin \theta~; \quad \bar{R}=\sqrt{\bar{\rho}^{2}+\bar{z}^{2}}~,\tan \theta=\frac{\bar{\rho}}{\bar{z}}~;
\nonumber
\\
\left(\frac{\partial}{\partial \bar{z}}\right)^{a}&=\cos \theta \left(\frac{\partial}{\partial \bar{R}}\right)^{a}-\frac{\sin \theta}{\bar{R}}\left(\frac{\partial}{\partial \theta}\right)^{a}~,
\\
\left(\frac{\partial}{\partial \bar{\rho}}\right)^{a}&=\sin \theta \left(\frac{\partial}{\partial \bar{R}}\right)^{a}+\frac{\cos \theta}{\bar{R}}\left(\frac{\partial}{\partial \theta}\right)^{a}~.
\end{align}
From the above coordinate transformation, one can construct the following vector field,
\begin{align}
\eta^{a}=\left(\frac{\partial}{\partial \bar{z}}\right)^{a}-i\left(\frac{\partial}{\partial \bar{\rho}}\right)^{a}
=e^{-i\theta}\left[\left(\frac{\partial}{\partial \bar{R}}\right)^{a}-\frac{i}{\bar{R}}\left(\frac{\partial}{\partial \theta}\right)^{a} \right]~,
\end{align}
which is null and is used to define the following scalar quantity, related to the $n$th order tensor multipole moment as, 
\begin{align}
f_{n}=\eta^{a_{1}}\cdots \eta^{a_{n}}P_{a_{1}\cdots a_{n}}~.
\end{align}
Since the Kerr-NUT spacetime is stationary and axi-symmetric, none of the tensors $P_{a_{1}\cdots a_{n}}$ can depend on the time or azimuthal coordinate $\phi$. Thus all these tensors are functions of $(\bar{R},\theta)$, which under the above coordinate transformation becomes a function of $\bar{z}$ and $\bar{\rho}$ respectively. Furthermore, the asymptotic point $P_{A}$ in the original Boyer-Lindquist coordinate system for the Kerr-NUT black hole spacetime correspond to $r\rightarrow \infty$. The same point in the unphysical metric corresponds to the limit, $\bar{R}\rightarrow 0$ and in the new $(\bar{z},\bar{\rho})$ coordinate system this translates into $\bar{z}^{2}+\bar{\rho}^{2} \rightarrow 0$. Often it is instructive to replace the above limit by $\bar{z}\rightarrow R_{c}$ and $\bar{\rho}\rightarrow iR_{c}$ respectively, where $R_{c}$ is a constant and then taking the $R_{c}\rightarrow 0$ limit. In particular, it is useful to define $y_{n}(R_{c})\equiv f_{n}(\bar{z}\rightarrow R_{c},\bar{\rho}\rightarrow iR_{c})$. In terms of this newly defined scalar function $y_{n}(R_{c})$, the recursion relation, presented in \ref{tensor_recursion}, takes the following form \cite{Backdahl:2005be},
\begin{align}
y_{n}=y_{n-1}'-2(n-1)\kappa_{\rm A}'y_{n-1}-\frac{(n-1)(2n-3)}{2}\mathcal{M}(R_{c})y_{n-2}~,
\end{align}
where, the function $\mathcal{M}(R_{c})$ is defined as,
\begin{align}\label{def_eq_kappa}
\mathcal{M}(R_{c})=\beta_{\rm A}''-\beta_{\rm A}'^{2}+\frac{2}{R_{c}}\beta_{\rm A}'-\kappa_{\rm A}''+\kappa_{\rm A}'^{2}~.
\end{align}
Here `prime' denotes derivative with respect to $R_{c}$ and the $n$th order multipole moment is given by $m_{n}\equiv y_{n}(0)$. Further, the quantity $\beta_{\rm A}$ corresponds to $\beta_{\rm A}\equiv \beta(\bar{z}\rightarrow R_{c},\bar{\rho}\rightarrow iR_{c})$, where $\beta$ is defined in \ref{beta_def}. Similarly, $\kappa_{\rm A}=\kappa(\bar{z}\rightarrow R_{c},\bar{\rho}\rightarrow iR_{c})$, where $\kappa$ captures the ambiguities in the conformal factor. 

Exploiting these relations, one can construct a scalar function, whose derivatives yield the multipole moments. For this purpose, we choose a function $\kappa_{\rm A}(R_{c})$, such that $\mathcal{M}(R_{c})$ is set to zero. Thus we need to make one more conformal transformation, $\widetilde{\Omega}=e^{\kappa}\Omega$ and hence the potential also gets rescaled, $\widetilde{\Phi}=e^{-\kappa/2}\bar{\Phi}$, as well as the metric, $\widetilde{h}_{ab}=e^{2\kappa}\bar{h}_{ab}$. Then introducing a new coordinate $\zeta$, which in terms of $R_{c}$ takes the form,
\begin{align}\label{def_zeta}
\zeta(R_{c})=R_{c}\exp \left[\kappa_{\rm A}(R_{c})-\beta_{\rm A}(R_{c})\right]~,
\end{align}
such that the scalar potential becomes, $\widetilde{\Phi}_{\rm A}=\widetilde{\Phi}_{\rm A}(R_{c}(\zeta))$. The asymptotic limit corresponds to $R_{c}\rightarrow 0$ and hence we have $\zeta\rightarrow 0$ in that limit as well. Thus all the multipole moments are derived by taking successive derivatives of the potential with respect to $\zeta$ and then taking the $\zeta \rightarrow 0$ limit, which yields,
\begin{align}\label{def_mult_mom}
\mathbb{M}_{n}\equiv \frac{2^{n}n!}{(2n)!}m_{n}=\frac{2^{n}n!}{(2n)!}\frac{d^{n}\widetilde{\Phi}}{d\zeta^{n}}\Big\vert_{\zeta \rightarrow 0}~.
\end{align}
This is what we will derive next, in the context of Kerr-NUT black hole spacetime, yielding its multipole moments. It is worth emphasizing that the above definition of the multipole moment is to reconcile the results with that derived by Hansen for Kerr spaceime. It must also be mentioned at this outset, since the structure of the multipole moment depends on the asymptotic structure of the spacetime, rather than near horizon or, near singularity behaviour, singularity in the Kerr-NUT spacetime will not affect the computation presented here.

From the expression for the Ernst potential presented in \ref{ernst_pot} above, we arrive at the following expression for the potential $\bar{\Phi}$, in the unphysical metric $\bar{h}_{ab}$, in the Kerr-NUT black hole spacetime,
\begin{equation}
\bar{\Phi}=\frac{1}{\bar{R}}\left[\left(1-\frac{M^{2}-a^{2}+N^{2}}{4}\bar{R}^2\right)^{2}-a^2 \bar{R}^2 \sin^2\theta\right]^{1/4} \left[\dfrac{1-\lambda-i \omega}{1+\lambda+i \omega} \right]~.
\label{eq:tildephi}
\end{equation}
In arriving at the above expression we have used the expression for the conformal factor $\Omega$ as presented in \ref{con_fac}. On the other had, using the expressions for the norm $\lambda$ and twist potential $\omega$ for the Kerr-NUT black hole, from \ref{KN_norm} and \ref{eq:twist_pot} respectively, one obtains the following expression,
\begin{eqnarray}
\frac{1-\lambda-i\omega}{1+\lambda+i \omega}=\frac{2N^{2}+2aN\cos\theta+2Mr+2i\left(aM\cos\theta+NM-Nr\right)}{2\left(r^{2}+a^{2}\cos^2\theta\right)-2Mr
-2i\left(aM \cos\theta+NM-Nr\right)}~.
\end{eqnarray}
However, the above expression involving $\lambda$ and $\omega$ is in terms of the old radial coordinate $r$, which needs to be converted to the new radial coordinate $\bar{R}$, such that the potential $\bar{\Phi}$ reads,
\begin{equation}
\bar{\Phi}=\left(\frac{\mathcal{A}}{\mathcal{B}}\right)\left[\left(1-\dfrac{M^2-a^2+N^2}{4}\bar{R}^2 \right)^2-a^2 \bar{R}^2 \sin^2\theta\right]^{1/4}~,
\end{equation}
where $\mathcal{A}$ and $\mathcal{B}$ has the following expressions,
\begin{align}
\mathcal{A}&=N^2\bar{R}+aN\bar{R} \cos\theta+M\left[1+M \bar{R}+\frac{\left(M^2-a^2+N^2\right)}{4}\bar{R}^{2}\right]~,
\\
\mathcal{B}&=\left[1+M\bar{R}+\frac{\left(M^2-a^2+N^{2}\right)}{4}\bar{R}^{2}\right]^{2}
-M\bar{R}\left[1+M\bar{R}+\frac{\left(M^2-a^2+N^{2}\right)}{4}\bar{R}^{2}\right]
\nonumber 
\\
&\hskip 1 cm +a^{2}\bar{R}^{2}\cos^2\theta-iM\bar{R}^2\left(N+a \cos\theta\right)+iN\bar{R}\left[1+M \bar{R}+\frac{\left(M^2-a^2+N^2\right)}{4}\bar{R}^2 \right]~.
\end{align}
Even though the above expression for $\bar{\Phi}$ looks complicated, as the earlier discussion shows, we actually require the above potential in a certain limit, in which, as we will see the above expression will simplify considerably. This is facilitated by first introducing the new coordinates, $\bar{z}$ and $\bar{\rho}$, such that $\bar{z}=\bar{R} \cos\theta$ and $\bar{\rho}=\bar{R}\sin\theta$. The asymptotic limit to the point $P_{\rm A}$ can be taken by simply substituting, $\bar{z}\rightarrow R_{\rm c}$ and $\bar{\rho}\rightarrow iR_{\rm c}$, such that we obtain $R\rightarrow \sqrt{\bar{\rho}^{2}+\bar{Z}^{2}}=\sqrt{(iR_{\rm c})^2+(R_{\rm c})^2}=0$. Under this limiting procedure, the expression for $\bar{\Phi}$ can be simplified and we finally arrive at the following expression for the potential $\bar{\Phi}$, 
\begin{equation}
\bar{\Phi}_{\rm A}(R_{c})=\left(M-iN\right)\frac{1+iaR_{\rm c}}{\left(1+a^{2}R_{\rm c}^{2}\right)^{3/4}}~.
\end{equation}
The above expression for the potential $\bar{\Phi}_{\rm A}$ at the asymptotic point can also be written in terms of the coordinate $\zeta$, defined in \ref{def_zeta}, which in the context of Kerr-NUT black hole spacetime reads, 
\begin{equation}
\zeta=\frac{R_{\rm c}}{1-a^{2}R_{\rm c}^{2}}~.
\end{equation}
As we have mentioned earlier, in this context, the multipole moments can be derived using a single scalar function, which in the present context reads,
\begin{equation}
\widetilde{\Phi}_{\rm A}=\exp\left[-\frac{\kappa_{\rm A}}{2}\right]\bar{\Phi}_{\rm A}=\frac{M-iN}{\sqrt{1-2ia\rho}}~,
\end{equation}
where, the quantity $\kappa_{\rm A}$ is defined as,
\begin{equation}
\kappa_{\rm A}=-\ln\left[\frac{1-a^{2}R_{\rm c}^{2}}{1+a^{2}R_{\rm c}^{2}}\right]~.
\end{equation}
Note that $\kappa$ denotes the ambiguity in defining the multipole moments and is fixed by setting the function $\mathcal{M}(R_{c})$ to zero in \ref{def_eq_kappa}. Finally, the expression for the multipole moments can be derived in terms of recursive derivatives of the above scalar function $\widetilde{\Phi}_{\rm A}$ and taking a cue from \ref{def_mult_mom}, we obtain,
\begin{equation}\label{mult_mom_KerrNUT}
\mathbb{M}_{n}=(M-iN)(ia)^{n}~.
\end{equation}
Note that in the limit, $N=0=a$, we get back the result that for Schwarzschild black hole, its mass determines all the multipole moments. While for $N=0$, we get back the Geroch-Hansen multipole moments for the Kerr black hole. In addition, the above expression for the multipole moment also satisfies the duality symmetry of the Kerr-NUT spacetime, as $\mathbb{M}_{n}(N=0)\leftrightarrow \mathbb{M}_{n}(M=0)$, modulo a sign, as we make the transformation, $M\leftrightarrow iN$. Thus the duality symmetry is preserved even at the level of multipole moments of the \KN spacetime. This shows the correctness of the multipole moment expression derived in \ref{mult_mom_KerrNUT}. Further, the multipole moment presented in \ref{mult_mom_KerrNUT} can be decomposed into mass and spin multipole moments, i.e., $\mathbb{M}_{n}=M_{n}+iS_{n}$. Such that, $M_{0}=M$, $M_{1}=Na$, $M_{2}=-Ma^{2}$, $M_{3}=-Na^{3}$, as well as, $S_{0}=-N$, $S_{1}=Ma$, $S_{2}=Na^{2}$, $S_{3}=-Ma^{3}$ and so on.

Let us point out another very interesting feature of the multipole moment for the Kerr-NUT black hole spacetime. Generically, the real part of $\mathbb{M}_{n}$ provides the mass multipole moments and the imaginary part of $\mathbb{M}_{n}$ gives the current multipole moments. For Kerr black hole all the odd mass multipole moments and even current multipole moments identically vanish, while for Kerr-NUT black hole spacetime, due to the presence of the NUT charge all mass and current multipole moments are non-zero. As we will observe, this drastically modifies the implications of these multipole moments for gravitational wave astronomy. The existence of mass and spin multipole moments at all orders for \KN black hole spacetime has to do with the asymmetry of the \KN solution about the equatorial plane. This connection will become clearer as we discuss the possible implications of our results on gravitational wave astronomy and hence possible constraint on the NUT charge. 

In passing, we should point out that another attempt to obtain the multipolar structure of the Kerr-NUT spacetime was carried out in \cite{Manko:2006bf} (see also \cite{Manko:2005nm}). However, their study uses a different formalism than the Geroch-Hansen one, which has been employed here. Also their work never discusses about the existence of the twist potential, the asymptotically flat 3-geometry, or the duality symmetry associated with the Kerr-NUT spacetime. In particular, it was not realized that even for asymptotically non-flat spacetimes, such as Kerr-NUT, it is indeed possible to employ the Geroch-Hansen formalism to obtain the multipole moments, owing to asymptotic flatness of the three-geometry. We must emphasize that none of these interesting and important aspects of the Kerr-NUT spacetime have been realized before. Furthermore, the final expression for the multipole moments of the Kerr-NUT black hole spacetime as obtained in \cite{Manko:2006bf}, matches with our study modulo a `-ve' sign, which appears due to different metric convention. 

Before concluding this section, let us briefly comment on the possible connection of the multipole moments derived earlier in this section with the Thorne's multipole moment. The main hurdle lies in the result that in the presence of a NUT charge it is not possible to express the \KN metric as $\eta_{\mu \nu}$, asymptotically. Since the Thorne's approach in computing the multipole moments crucially hinges on the asymptotic behaviour of the metric, it needs to reconstructed from scratch. However, as we have demonstrated in \ref{AppThorne}, expansion of the $g_{tt}$ and the $g_{t\phi}$ components of the \KN metric involves both $\mathcal{O}(1/r)$ and $\mathcal{O}(1/r^{2})$ terms. This is unlike the situation for the Kerr black hole, where $g_{tt}$ does not involve $\mathcal{O}(1/r^{2})$ term and $g_{t\phi}$ does not involve $\mathcal{O}(1/r)$ term. Thus, for the Kerr-NUT black hole spacetime, both the mass and spin monopole moments exist and they can be derived from the $\mathcal{O}(1/r)$ term of the $g_{tt}$ and the $g_{t\phi}$ components respectively. As one can immediately verify from \ref{AppThorne}, the monopole moments are proportional to the mass $M$ and the NUT charge $N$ respectively. Similarly, there will be both mass and spin dipole moments proportional to $Ma$ and $Na$ respectively, see \ref{AppThorne}. These results are consistent with the Geroch-Hansen formalism, as we can see by substituting $n=0$ and $n=1$ in       \ref{mult_mom_KerrNUT}. Therefore, the Geroch-Hansen and the Thorne's approach provide identical expression for the leading order multipole moments of the \KN black hole. To see the equivalence in general, one needs to modify the Thorne's approach by taking into account the asymptotic structure of the \KN spacetime, which we leave for the future. This concludes our derivation of the multipole moments for the Kerr-NUT spacetime, which we will apply in the next section to understand possible implication of the NUT charge in the gravitational wave astronomy.  

\section{Gravitational wave observables in terms of the multipole moments of the Kerr-NUT spacetime}\label{KN_GW}

The connection between the multipole moments and the gravitational wave observables is a very intriguing one, since it extends both ways. On one side, knowing the multipole moments of a compact object one can predict the observables associated with gravitational waves, while on the other, the gravitational wave observables can tell us about the multipole moments of a compact object. The second approach is more useful and often tries to express the gravitational wave observables in terms of the multipole moments of the gravitational field. Since, we have derived all the multipole moments of the Kerr-NUT spacetime, it is important to ask, if we can get a handle on the numerical estimation of the NUT charge \'{a} la gravitational wave observations. To understand this, one may follow the approach presented in \cite{Ryan:1995wh}, with two major differences. As we will demonstrate, these two issues will be sufficient enough to put some constraint on the NUT charge using the geometry of the Kerr-NUT spacetime. 

It is worthwhile to emphasize another interesting point regarding the current analysis at this stage. The computation involving gravitational wave astronomy, as well as computation of multipole moment requires asymptotic flatness, which at first sight does not seem to hold good in Kerr-NUT spacetime. However, it is only the asymptotic flatness on the induced metric on a $t=\textrm{constant}$ slice, which matters for the multipole moment computation and also for the gravitational wave astronomy, since it is only the spatial part of the metric tensor which carries the dynamical degrees of freedom for the gravitational field. Thus following \ref{lower_dim_KN}, it is clear that the spatial sector of the metric, or the induced metric on a $t=\textrm{constant}$ Cauchy slice, is indeed asymptotically flat. Thus the gravitational degrees of freedom indeed propagates in an asymptotically flat spacetime and hence can be used in the study of the gravitational wave astronomy.

The analysis relating gravitational wave observables with multipole moments serves several purposes, first of all it gives an idea about the first few multipole moment of the compact object. More so, it provides a direct hint to the validity of the no-hair theorem in case the compact object is a black hole. However, most of the analysis in this direction bears two crucial assumptions --- (a) the gravitational field produced by the compact object is reflection symmetric about the equatorial plane and as a consequence odd mass multipole moments and even current multipole moments vanish and (b) the in-spiral phase can be approximated to be a circular orbit on the equatorial plane. In addition there are other assumptions involving geodesic orbits, adiabatic evolution, ignoring back-reaction problem etc. Even though most of these assumptions are valid for Kerr-NUT spacetime, the two assumptions pointed out above does not hold. Namely, the \KN spacetime is \emph{not} reflection symmetric about the equatorial plane, as one can immediately check by substituting $\theta\rightarrow (\pi/2)+\theta$ in \ref{metric_KN_old}. As a consequence there are odd mass multipole moments and even current multipole moments in a \KN spacetime, see \ref{mult_mom_KerrNUT}. This is a distinct signature of the NUT charge, which must be looked for in the gravitational wave observables. As a consequence the recursion relation, connecting multipole moments with gravitational wave observables get modified. In addition, for \KN black hole spacetime there exist \emph{no} circular orbits on the equatorial plane, again due to the very existence of NUT charge. As we will see this can severely constrain the parameter space for the NUT charge if we want the assumptions coined before to hold true. 
\subsection{Connecting multipole moments with observables: Modified recursion relation}

In this section, we will provide the connection between the multipole moment structure derived in the previous section with the gravitational wave observables, which we will now introduce. This will provide an interesting avenue to look for the NUT charge from gravitational wave observations, through the multipolar structure of the central massive object. We will first present these observables, solely in the context of the Kerr-NUT geometry, before commenting on the general structure through a modified recursion relation. 

We will provide three such observables, which are intimately connected with the gravitational wave emission from the inspiral of a compact object onto the central massive object, which could be a Kerr-NUT black hole. During the early phase of the inspiral, we can approximate the in-falling object as one moving on an almost circular orbit, with its radius decreasing gradually. In the absence of tidal heating, the loss of energy due to the emission of gravitational waves is solely determined by the rate of change of conserved energy as the radius of the circular orbit decreases. This is best demonstrated by the following quantity \cite{Ryan:1995wh}, 
\begin{equation}\label{grav_energy}
\Delta E_{\rm GW}=-\Omega_{\rm circ} \frac{dE_{\rm circ}}{d\Omega_{\rm circ}}~,\quad \frac{E_{\rm circ}}{m}=\frac{-g_{tt}-g_{t\phi}\Omega_{\rm circ}}{\sqrt{-g_{tt}-2g_{t\phi}\Omega_{\rm circ}-g_{\phi \phi}\Omega_{\rm circ}^{2}}}~,
\end{equation}
where, $E_{\rm circ}$ is the energy of the inspiralling compact object with mass $m$ on a circular orbit, $\Omega_{\rm circ}=(d\phi/dt)$ is the angular frequency associated with the circular orbit and $\Delta E_{\rm GW}$ denotes the energy radiated away by the gravitational waves. It must be noted that the use of $\Delta$ in the above expression is to quantify the change in energy through gravitational waves and has no connection with the $g_{rr}$ metric element of the Kerr-NUT spacetime.

The other two observables are the precession frequencies associated with --- (a) departure of the orbit from being circular and (b) departure from the equatorial plane. Both of these frequencies will influence the spectrum of the emitted gravitational waves and hence will have direct observable consequences. In terms of the metric elements and the angular frequency of the circular orbit $\Omega_{\rm circ}$, the precession frequencies can be expressed as, 
\begin{align}
\Omega_{p}&=\Omega-\bigg\{-\frac{g^{pp}}{2}\bigg[\left(g_{tt}+g_{t\phi}\Omega\right)^{2}\partial_{p}^{2}\left(\frac{g_{\phi \phi}}{\rho^{2}}\right)
\nonumber
\\
&\hskip 1cm -2\left(g_{tt}+g_{t\phi}\Omega\right)\left(g_{t\phi}+g_{\phi\phi}\Omega\right)\partial_{p}^{2}\left(\frac{g_{t\phi}}{\rho^{2}}\right)+\left(g_{t\phi}+g_{\phi\phi}\Omega\right)^{2}\partial_{p}^{2}\left(\frac{g_{tt}}{\rho^{2}}\right) \bigg] \bigg\}^{1/2}~,
\end{align}
where, $p=(\rho,z)$ or $(r,\theta)$, depending on whether we are using cylindrical coordinate system or Boyer-Lindquist coordinate system, respectively. Note that the terms involving single derivative with respect to $\rho$ and $z$ as well as with respect to $r$ and $\theta$ are absent. This is due to the $\rho\rightarrow -\rho$ symmetry, or in other words, $\theta \rightarrow -\theta$ symmetry of the \KN spacetime. It is worth emphasizing that these frequencies make sense only for small perturbations around the equatorial plane, i.e., for small values of $z$ or, of $[\theta-(\pi/2)]$, respectively.

It is customary to express the above three observables --- (a) energy lost due to gravitational radiation $\Delta E_{\rm GW}$, (b) precession frequencies $\Omega_{r}$ and (c) $\Omega_{\theta}$, as a power series in the velocity of the compact object inspiraling the central \KN black hole. This is considered as a post-Newtonian (henceforth referred to as PN) expansion of these observables and can be achieved along the following lines. First of all, the metric functions and their derivatives can be expanded as a power series in $(1/r)$ and as a consequence, the angular frequency $\Omega_{\rm circ}$ can also be expressed as a power series in $(1/r)$, such that $\Omega_{\rm circ}=\sqrt{(M/r^{3})}[1+\mathcal{O}(r^{-1/2})]$. This can be inverted, yielding $r$ as a function of the angular frequency $\Omega_{\rm circ}$. Therefore, using the expansion of all the observables, namely $\Delta E_{\rm GW}$, $\Omega_{r}$ and $\Omega_{\theta}$, in powers of $(1/r)$, one can express these observables as a power series in the angular frequency $\Omega_{\rm circ}$. This in turn can be converted to a power series in the velocity of the inspiraling object, as $\Omega_{\rm circ} \sim v^{3}$ and hence the desired PN expansion of these observables can be obtained. 

In the PN expansion of the gravitational wave observables, each coefficients of the expansion depend crucially on the multipole moments of the central massive object, which the other compact object is inspiraling. This expansion modifies significantly as the spacetime described by the central massive object does not have reflection symmetry about the equatorial plane, as we will demonstrate below by taking the central massive object to be described by the Kerr-NUT geometry.  

Using the metric elements of the \KN spacetime described in \ref{metric_KN_old}, the angular velocity of the circular orbit on the equatorial plane can be immediately determined and hence we can express the radial coordinate $r$ in the Boyer-Lindquist coordinate system as a power series in $\Omega$, which takes the following form,
\begin{equation}
r=\left(\dfrac{M}{\Omega_{\rm circ}^2}\right)^{1/3}\left[1+\left(\dfrac{2N^2}{3 M^{4/3}} \right)\Omega_{\rm circ}^{2/3}+\left(-\dfrac{2a}{3}\right)\Omega_{\rm circ}+\mathcal{O}(\Omega_{\rm circ}^{4/3})\right]~.
\label{eq:angular_fre_NUT}
\end{equation}
Here, the first term yields the Keplarian contribution and the other terms in the above expansion arise due to the presence of the rotation and the NUT charge. Interestingly, the NUT charge appears at a lower order than the rotation and hence it seemingly provides a larger contribution to the angular velocity. However, as we will see later that is not the case. Thus using the above expansion of the radial coordinate in terms of the angular velocity, we obtain the following PN expansion for the observables $\Delta E_{\rm GW}$, $\Omega_{r}$ and $\Omega_{\theta}$,
\begin{align}
\Delta E_{\rm GW}&=\dfrac{v^2}{3}+\left(-\dfrac{1}{2}-\dfrac{4(-N)^2}{9M^2} \right)v^4+\dfrac{20(aM)}{9M^{2}}v^5
-\Bigl(\frac{27}{8}-\frac{(-Ma^2)}{M^3}+\frac{2(-N)^2}{M^2}\Bigr)v^6+\mathcal{O}(v^7)~,
\\
\dfrac{\Omega_r}{\Omega}&=\left(3+\dfrac{(-N)^2}{M^2}\right)v^2-\dfrac{4(aM)}{M^{2}}v^{3}
+\Bigl(\dfrac{9}{2}-\dfrac{19 (-N)^4}{ M^4}-\dfrac{3(-Ma^{2})}{2M^3}\Bigr)v^{4}
\nonumber
\\
&\hskip 1 cm -\Bigl(-\dfrac{10(aN)(-N)}{3M^3}+\dfrac{10(aM)}{M^{2}}\Bigr)v^5+\mathcal{O}(v^6)~, 
\\
\dfrac{\Omega_{\theta}}{\Omega}&=-\dfrac{2(-N)^2}{M^2}v^2+\dfrac{2(aM)}{M^{2}}v^3
+\left(\dfrac{22(-N)^4}{3M^4}+\dfrac{3(-Ma^2)}{2M^3}+\dfrac{8(-N)^2}{M^2}\right)v^4
\nonumber
\\
&\hskip 1 cm +\dfrac{22(aN)(-N)}{3M^3}v^5+\mathcal{O}(v^6).
\end{align}
In the above expression, interestingly, the contribution from the NUT charge appears prior to the angular momentum of the black hole for all the three observables. For example, in the expression for energy radiated away by gravitational waves, the NUT charge contributes in 2PN order, while the angular momentum starts contributing from the 2.5PN term. This suggest that the NUT charge will contribute at a leading order compared to the angular momentum of the black hole and hence can be used to provide stringent constraint on the NUT parameter. In particular, the 2PN and the 2.5PN terms will become comparable, if the NUT charge and the rotation parameter, satisfies the following relation, $(N/M)^{2}\sim (a/M)v$. As we will show later, the ratio $(N/M)$ is always much smaller, compared to $(a/M)$ and hence the effect of rotation is always larger. 

These expressions are in terms of the hairs of the Kerr-NUT black hole spacetime. However, the hairs can be identified with various orders of mass and spin multipole moments of the Kerr-NUT black hole spacetime, starting from \ref{mult_mom_KerrNUT}. This provides the general PN expansion of the observables associated with the gravitational wave emission from the inspiral of a compact object around a central massive object, which is not reflection symmetric about the equatorial plane as,
\begin{align}
\Delta E_{\rm GW}&=\dfrac{v^2}{3}+\left(-\dfrac{1}{2}-\dfrac{4S_{0}^2}{9M^2} \right)v^4+\dfrac{20S_{1}}{9M^{2}}v^5
+\Bigl(-\frac{27}{8}+\frac{M_{2}}{M^3}-\frac{2S_{0}^2}{M^2}\Bigr)v^6+\mathcal{O}(v^7)~,
\\
\dfrac{\Omega_r}{\Omega}&=\left(3+\dfrac{S_{0}^2}{M^2}\right)v^2-\dfrac{4S_{1}}{M^{2}}v^{3}
+\Bigl(\dfrac{9}{2}-\dfrac{19S_{0}^4}{M^4}-\dfrac{3M_{2}}{2M^3}\Bigr)v^{4}
\nonumber
\\
&\hskip 1 cm +\Bigl(\dfrac{10S_{0}M_{1}}{3M^3}-\dfrac{10S_{1}}{M^{2}}\Bigr)v^5+\mathcal{O}(v^6)~, 
\\
\dfrac{\Omega_{\theta}}{\Omega}&=-\dfrac{2S_{0}^2}{M^2}v^2+\dfrac{2S_{1}}{M^{2}}v^3
+\left(\dfrac{22S_{0}^4}{3M^4}+\dfrac{3M_{2}}{2M^3}+\dfrac{8S_{0}^2}{M^2}\right)v^4
\nonumber
\\
&\hskip 1 cm +\dfrac{22S_{0}M_{1}}{3M^3}v^5+\mathcal{O}(v^6)~.
\end{align}
Note that in the above expansion, both the even and odd orders of mass and spin multipole moments are present. In addition, if we set the odd mass multipole moments and even spin multipole moments to be vanishing, we get back the results presented in \cite{Ryan:1995wh}. Thus the above PN expansion of the gravitational wave observables generalizes the earlier approaches significantly, by incorporating mass and spin multipole moments of all orders. Furthermore, the results presented above is applicable for generic spacetime geometry, which may or may not admit reflection symmetry about the equatorial plane.  

The above describes the observables associated with the gravitational waves in the context of the Kerr-NUT black hole spacetime. As we have observed, breaking of reflection symmetry has significant effect on the gravitational wave observables. There are additional multipole moments in these contexts, which can affect these observables significantly. In what follows we will briefly describe the strategy one may follow in a generic context of which \KN spacetime is just a special case, where both mass and current multipole moments have even as well as odd sectors. To elaborate on this, we note that asymptotically, in the limit of large $r$, the conformal factor takes the following form, $\Omega=r^{-2}$, see \ref{coord_transform} and  \ref{con_fac}. In this limit, the cylindrical coordinates $\rho$ and $z$ introduced in \ref{KN_Cylindrical} can be expressed as, $\rho=r\sin \theta$ and $z=r\cos \theta$, such that $\Omega=(\rho^{2}+z^{2})^{-1}$. Thus the Ernst potential, defined in \ref{ernst_pot_mod}, can be expressed in terms of the unphysical potential $\bar{\Phi}$, in the following form,
\begin{equation}
\varepsilon=\frac{\sqrt{\rho^{2}+z^{2}}-\bar{\Phi}}{\sqrt{\rho^{2}+z^{2}}+\bar{\Phi}}~.
\end{equation}
As we have demonstrated earlier, various derivatives of the function $\bar{\Phi}$ yields the multipole moments of various orders. Also the stationarity and axi-symmetry of the problem suggests that $\bar{\Phi}$ does not depend on $t$ and $\phi$, while it depends only on the coordinates $\rho$ and $z$. Since the multipole moments are coefficients of various powers of the radial coordinate, we may expand the unphysical potential $\bar{\Phi}$ as,
\begin{equation}\label{expansion}
\bar{\Phi}=\sum_{j,k}a_{jk}\frac{\rho^{j}z^{k}}{\left(\rho^{2}+z^{2}\right)^{j+k}}~, 
\end{equation}
where $j$ and $k$ are both non-negative. Since, $\theta\rightarrow -\theta$ is a symmetry of the \KN spacetime, as evident from the metric depicted in \ref{metric_KN_old}, the above expansion should also be invariant under $\rho\rightarrow -\rho$. Since under the above transformation of the angular coordinate $\theta$, $\rho\rightarrow -\rho$ and $z\rightarrow z$, respectively. Thus the index $j$ should take only even values. Since, the Kerr-NUT spacetime has no reflection symmetry about the equatorial plane, as described before, the coefficients $a_{jk}$ have both real and imaginary parts for even as well as odd values of the index $k$. This is in striking contrast with the Kerr spacetime, where the coefficients $a_{jk}$ are real for even $k$ and are purely imaginary for odd $k$. As we will see, this will have major impact on the dependence of the gravitational wave observables on the multipolar structure of the Kerr-NUT spacetime. 

Since, we know the metric elements near and on the equatorial plane, it is expected that the coefficients $a_{j0}$ and $a_{j1}$ are well known. Given these coefficients, the rest of them can be determined using a recursion relation. For an even integer $m$, we assume that all the $a_{j0}$ coefficients for $j=0,2,\ldots m$ and all the $a_{j1}$ coefficients for $j=0,2,\ldots (m-2)$ are known. While for an odd integer m, all the $a_{j0}$ and $a_{j1}$ coefficients with $j=0,2,\ldots (m-1)$ are known. Then all other coefficients $a_{jk}$, for $j+k\leq m$ can be determined from the following recursion relation (for a derivation, see \ref{AppRec})
\begin{align}
a_{r,s+2}&=\frac{1}{(s+1)(s+2)}\Bigg\{-\left(r+2\right)^{2}a_{r+2,s}+\sum a_{kl}a^{*}_{r-k-p,s-l-q}\Bigl[a_{p+2,q-2}(p+2)(p+2-2k)
\nonumber 
\\
&\hskip 3 cm +a_{p-2,q+2}(q+2)(q+1-2l)+a_{pq}(p^2+q^2-4p-5q-2pk-2ql-2)\Bigr] \Bigg\}~.
\end{align}
Since the symmetry of the metric under the transformation $\theta\rightarrow -\theta$ must be respected, the integers $r$, $k$ and $p$ must be even. In addition, we must have the following restrictions, $0\leq k\leq r$, $0\leq l \leq (s+1)$, $0\leq p \leq (r-k)$ and $-1\leq q \leq (s-l)$. Using these restrictions, it follows that the above recursion relation demands expressibility of $a_{r,s+2}$ in terms of all the $a_{j0}$ and $a_{j-1,1}$ coefficients respectively. 

The determination of the coefficients $a_{j0}$ and $a_{j1}$ follows from the observables $\Delta E_{\rm GW}$, $\Omega_{\rho}$ and $\Omega_{z}$ respectively. However, unlike the reflection symmetric case, here all the coefficients will have both real and imaginary parts. Then the multipole moments can be related to these coefficients by various derivatives of the unphysical potential $\bar{\Phi}$. Here also we have,
\begin{align}
a_{0\ell}=M_{\ell}+iS_{\ell}+\textrm{lower~order~moments}~.
\end{align}
Note that in the reflection symmetric case, $a_{0\ell}$ is real for even $\ell$ and is purely imaginary for odd $\ell$. While in the present context $a_{0\ell}$ has both real and imaginary parts for all possible values of $\ell$ and is consistent with the earlier discussions. Thus the mass and spin multipole moments exist at all orders. This finishes our discussion of relating the gravitational wave observables with multipole moments of the central massive object, which is not symmetric about the equatorial plane, for an binary of incomparable masses. We will now describe the constraints on the NUT charge using the non-existence of circular orbits on the equatorial plane in the presence of NUT charge. 

\subsection{Circular orbits on the equatorial plane: Constraints on the NUT charge}

The formalism presented in the previous section helped us to obtain the energy radiated in the form of gravitational waves as well as the precession frequencies in the case of an incomparable binary. As we have emphasized, there are significant departures of these results from the standard expectations (with the case of the Kerr black hole at the back of the mind) due to the presence of the NUT charge. This is because, in the presence of the NUT charge, the mass and the spin multipole moments of all orders are present. This leads to additional terms in the gravitational wave observables, e.g., energy radiated by the binary system, which have direct observable consequences. However one crucial assumption that has gone into the above computation, is that the orbits are (nearly) circular and confined on the equatorial plane. This assumption is easily seen to be valid in the case of Kerr spacetime; however, for the Kerr-NUT geometry this is no longer true. In particular, one can demonstrate that for generic initial data, there are no circular timelike geodesics that can exist on the equatorial plane \cite{Jefremov:2016dpi,Mukherjee:2018dmm}. Therefore, the computation presented above, without addressing this subtlety would be erroneous, which we aim to address in this section.

Despite having several intriguing properties, thereby modifying the multipole moments in a non-trivial manner, the presence of the NUT charge also results into a problematic feature regarding the orbital dynamics on the equatorial plane. As it was pointed out in Ref. \cite{Jefremov:2016dpi}, and later in Ref. \cite{Mukherjee:2018dmm}, there exists no stable circular timelike geodesic on the equatorial plane of a \KN spacetime. This becomes a major hindrance to extend the formalism presented in \cite{Ryan:1995wh}, as it had been built upon the assumption of existence of circular orbits on the equatorial plane. This formalism proposed an unique mechanism to relate the multipole moments with energy radiation and precession frequencies --- making it an important tool to investigate evolution of a binary especially discussing extreme mass ratio inspiral (EMRI). However, the vast assumptions of equatorial and circular geodesics makes this formalism somewhat restricted to realistic astrophysical events. 

To understand the problem associated with the existence of circular orbits on the equatorial plane, let us start by assuming that such a circular orbit does exist. This implies that the trajectory is given by $r=r_{\rm c}$ and $\theta=(\pi/2)$, where $r_{\rm c}$ is the radius of the circular orbit on the equatorial plane. For the circular orbit to continue remain circular on the equatorial plane, it is necessary that $\dot{r}=\ddot{r}=0$ as well as $\dot{\theta}=0=\ddot{\theta}$. Here \enquote*{dot} denotes the differentiation of the geometrical quantity, with respect to the proper time $\tau$. For Kerr black hole one can immediately check that these conditions are identically satisfied. On the other hand, in presence of the NUT charge, the condition $\dot{\theta}=0$ determines the Carter constant to be non-zero and equal to $N^2$; thus following \cite{Mukherjee:2018dmm} one can immediately demonstrate that $\ddot{\theta} \neq 0$ (see \ref{AppCirc} for a detailed derivation). This implies that, even if one starts with a planner orbit, the particle will eventually move out of the equatorial plane at a later instant of time in the presence of the NUT charge. Therefore, timelike circular orbits are unlikely to appear on the equatorial plane of the Kerr-NUT black hole. Even then, the calculation presented above makes complete sense. This is because, there are two time scales associated with this problem --- (a) the time in which the particle goes sufficiently away from the equatorial plane and (b) the time scale over which the loss due to gravitational radiation is significant. If the particle remains confined to the equatorial plane or to its immediate vicinity for an interval of time, which is sufficiently long for the amount of emitted gravitational radiation to be significant, the above analysis will be directly applicable. As we demonstrate below, for small enough NUT charge (more quantitative estimation of the smallness of the NUT charge will be provided shortly) such a scenario indeed exists.

\begin{figure}[htp]
\subfloat[The above figure demonstrates the angular variation of the circular orbit around the equatorial plane orbiting a Kerr-NUT black hole with NUT charge, $N=0.01M$.\label{Figure_01A}]{%
\includegraphics[height=5.5cm,width=.49\linewidth]{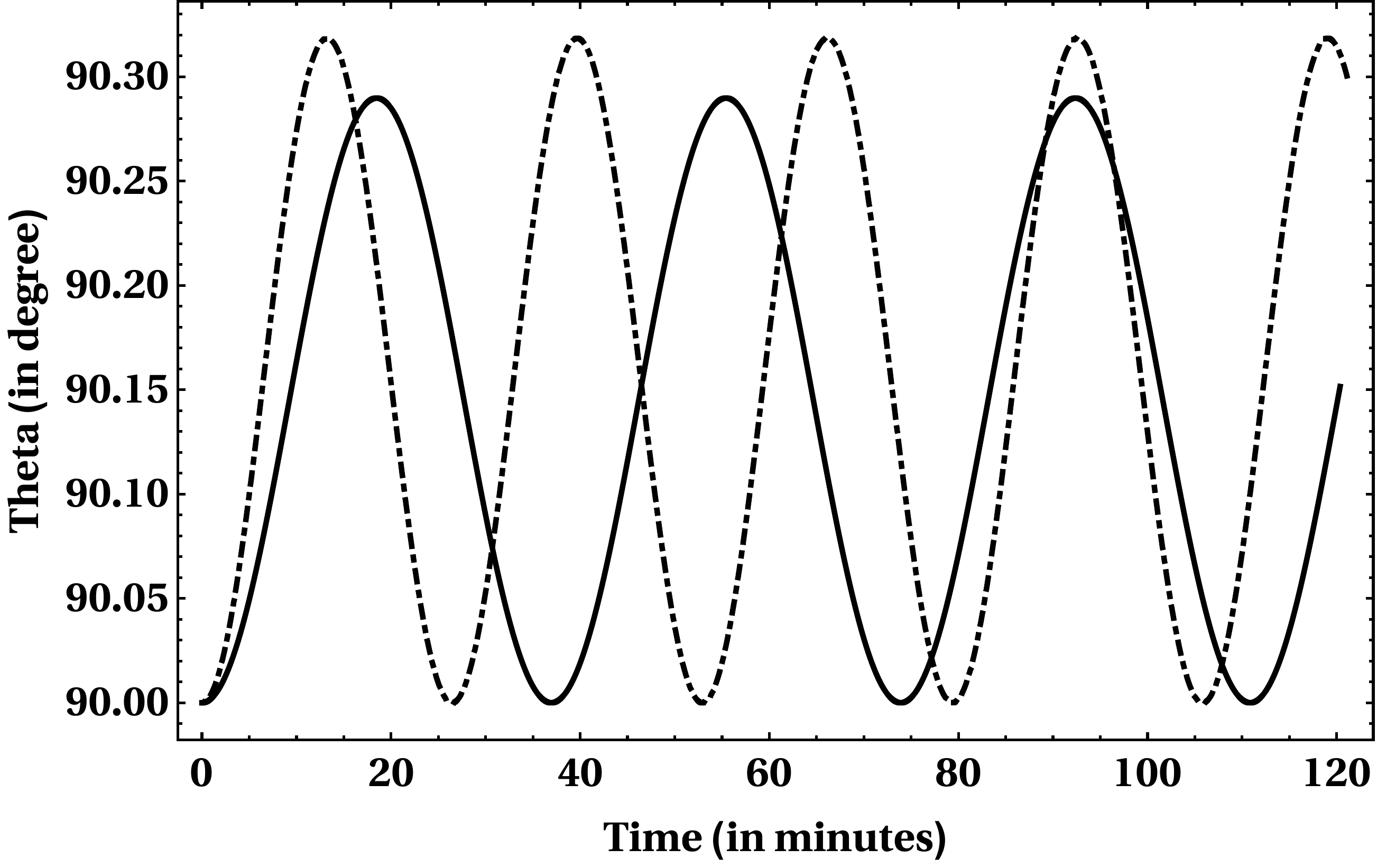}%
}\hfill
\subfloat[The deviation of the circular orbit from the equatorial plane has been depicted, as the NUT charge of the central massive object is taken to be $N=0.1M$. \label{Figure_01B}]{%
\includegraphics[height=5.5cm,width=.49\linewidth]{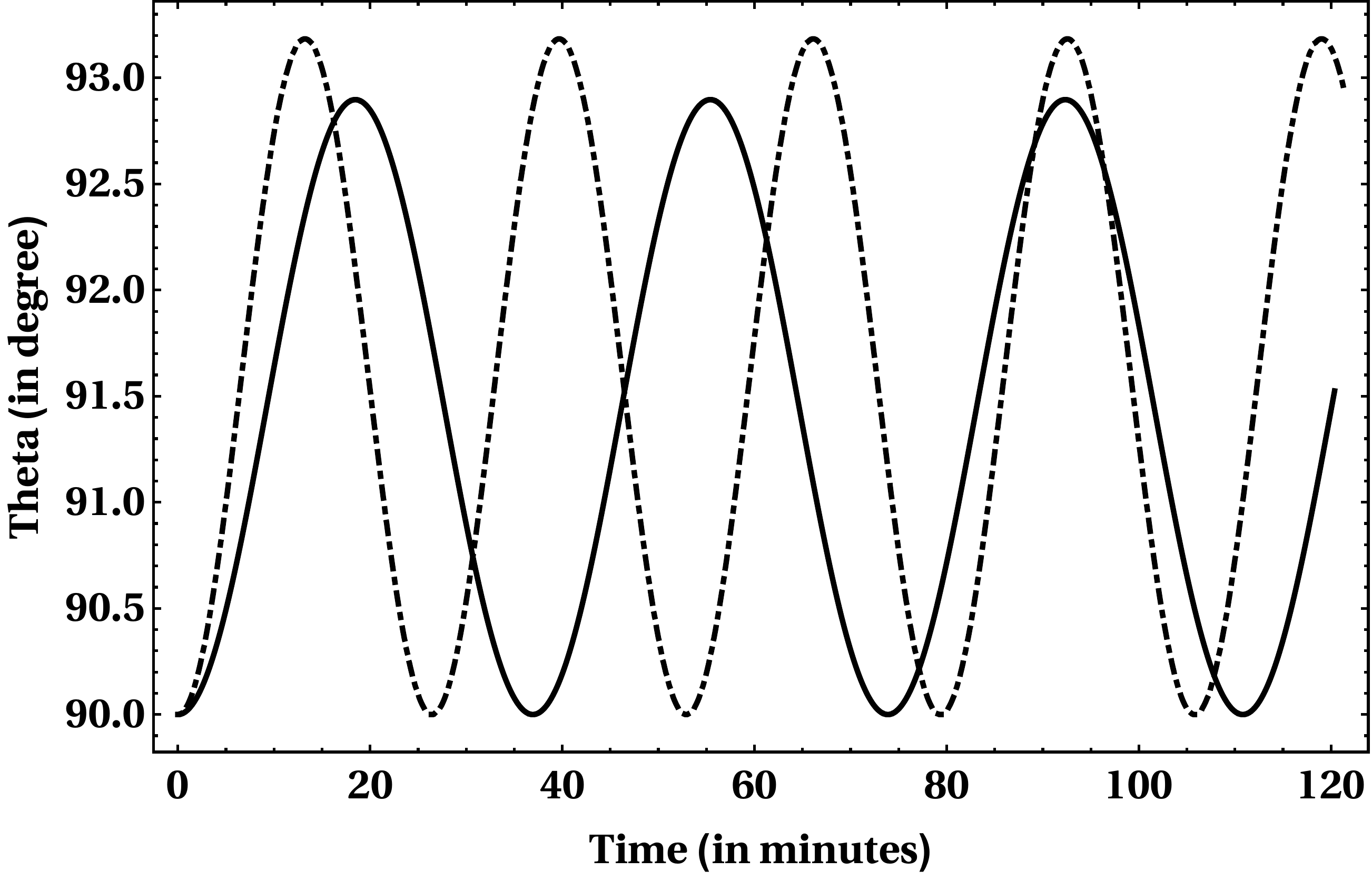}%
}\\
\hfill
\centering
\subfloat[The deviation from the equatorial plane is presented for the case $N=M$. \label{Figure_01C}]{%
\includegraphics[height=5.5cm,width=.49\linewidth]{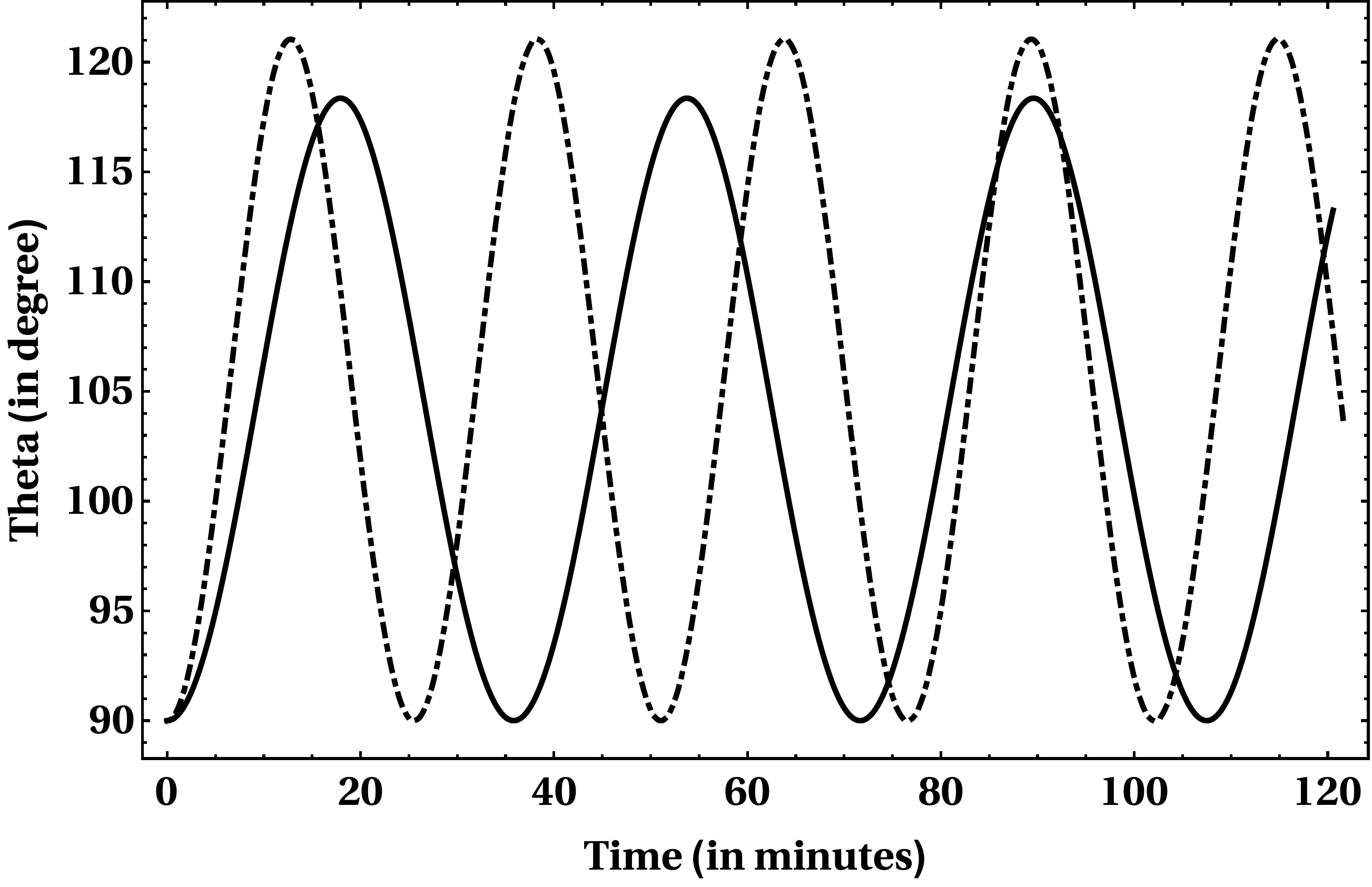}%
}
\caption{The above figures capture the angular deviation of the orbit from the equatorial plane for various values of the NUT parameter and for different radii of the circular orbit. For the solid curve in the above figures, the radius of the circular orbit $r_{\rm c}$ is taken to be $50M$, while for the dashed curve $r_{\rm c}$ is $40 M$. Given that the NUT charge and the radius of the orbit are responsible for this off-equatorial plane motion --- lesser the NUT charge and larger the radius, lesser is the angular deviation. See text for more discussion.}
\label{Figure_01}
\end{figure}

Let us consider an example in which a particle starts orbiting the central massive object in a circular trajectory on the equatorial plane. Assuming that the geometry of the central object is described by the Kerr-NUT spacetime, we set the Carter constant to be, $\lambda=N^2$; therefore, according to the previous discussion suggests, $\dot{\theta}$ identically vanishes, but $\ddot{\theta}$ does have a non-zero value (see, e.g., \ref{AppCirc}). A nonzero $\ddot{\theta}$ would change $\dot{\theta}$, and then $\theta$ --- finally, the particle will move away from the equatorial plane. The question is, how quick is this process involving departure from the equatorial plane. To study the evolution of the particle as it starts on the equatorial plane, we employ the Euler method \cite{ferziger1981numerical} to integrate the geodesic equations in the Kerr-NUT black hole spacetime. With the standard expressions for $\dot{\theta}$ and $\ddot{\theta}$, to leading order we may assume the evolution equations are given as follows:
\begin{equation}\label{iteration}
\dot{\theta}_{\rm new}=\dot{\theta}_{\rm old}+(d^{2}\theta/d\tau^{2})_{\rm old}~ \delta \tau~, \quad 
\theta_{\rm new}=\theta_{\rm old}+(d\theta/d\tau)_{\rm old}~ \delta \tau~, ~\text{and} \quad t_{\rm new}=t_{\rm old}+\mathcal{U}^{t} \delta \tau~,
\end{equation}
where, $\mathcal{U}^{t}$ is the t-component of the 4-velocity of the orbiting particle. This can be written in terms of the energy $E$ and the momentum $L_{\rm z}$ as follows,
\begin{equation}
\mathcal{U}^{t}=-g^{tt}E+g^{t\phi}L_{\rm z}~,
\end{equation}
where $g^{tt}$ and $g^{t\phi}$ are the metric components of the Kerr-NUT black hole spacetime. The other notations, such as \enquote*{old} and \enquote*{new}, introduced in \ref{iteration} is given for past and future step differing by the proper time interval $\delta \tau$, respectively, while iterating for the numerical analysis. We assume that the particle starts at $\tau=0$, and the circular orbit can have a radius of $r_{\rm c}\sim \mathcal{O}(50M)$. Using these initial conditions, we determine the orbit at a later instant by solving the geodesic equations, numerically. The outcome of such a numerical analysis has been depicted in \ref{Figure_01}, which shows an interesting behaviour. As evident, the orbits slowly evolve away from the equatorial plane and starts to oscillate about another planar section, different from $\theta=(\pi/2)$. Interestingly, this oscillation does not decay in time, rather the orbit keeps oscillating with a constant amplitude. The amplitude depends crucially on the value of the NUT parameter, and for larger values of the NUT charge the amplitude increases, thereby affecting the deviation from the equatorial plane. For example, in \ref{Figure_01A}, the deviation from the equatorial plane has been plotted for $N=10^{-2}M$, and as one can clearly observe, the angular deviation becomes $\Delta \theta \approx 0.3^{\circ}$. On the other hand, for the case $N=M$, given in \ref{Figure_01C}, the deviation from the equatorial plane becomes large, as $\Delta \theta \approx 20^{\circ}$. Therefore, for a sufficiently small value of the NUT parameter, possibly with $(N/M)<10^{-2}$, the formalism developed in the earlier section will be highly appropriate. Thus there exist a range of the NUT charge, for which the particle almost remains confined to the equatorial plane and hence the earlier discussion comes to life. The other important factor which also contributes to the deviation of the orbit from the equatorial plane, is the radius of the circular orbit. As all the plots in \ref{Figure_01} demonstrates, for a larger radius, the angular deviation is smaller compared to an orbit of smaller radius. This can also be understood from \ref{eq:angular_fre_NUT}, which relates the angular frequency with the radius of the circular orbit and the NUT charge. Since the orbits with larger radius describes the early phase of a binary inspiral, while smaller radius corresponds to a binary at the late stage of the inspiral, we can safely state that the formalism presented here is suitable to describe the early phase of the evolution of a binary system, with the massive object described by the Kerr-NUT geometry with $(N/M)< \mathcal{O}(10^{-2})$. 

This shows that even though in the PN expansion of the observables, the NUT charge appears at a lower PN order than the rotation parameter, it is the rotation parameter which will dominate the picture. As emphasized earlier, the contribution from the NUT charge and the rotation parameter will be comparable if $(N/M)^{2}\sim (a/M)v$. Thus for the typical choice of the NUT charge, consistent with the quasi-circular orbit on the equatorial plane, we have $(N/M)\sim 10^{-3}$. A typical rotation parameter will have the following estimation, $(a/M)\sim 0.1$, such that the effect from the NUT charge and rotation will be comparable for $v\sim 10^{-5}$. This corresponds to typical velocity of an inspiraling system and hence for all practical purposes, the NUT charge and the rotation parameter contributes equally to the gravitational radiation. An identical consideration applies to other observables as well. Therefore, a proper analysis of the inspiral part of the merger events seen in Advanced LIGO experiment can be used to provide bound on the NUT charge, which we leave for the future. 

\section{Conclusion}

In this work, we have discussed the multipolar structure of the \KN spacetime, which have enabled us to address some of the theoretical and observational implications pertaining to it. Given the not-so-obvious result that, asymptotically the Kerr-NUT spacetime does not reduce to flat spacetime metric $\eta_{\mu \nu}$, one may argue that the \GH~formalism is probably not applicable to derive the multipole moments of \KN spacetime. However, as discussed in \ref{Sec_KN} (see also \cite{Misner:1963fr}), the asymptotic limit of the \KN spacetime is non-trivial, since even though the \KN spacetime does not reduce to flat spacetime metric, the spacetime curvature vanishes asymptotically $\sim \mathcal{O}(1/r^3)$. Due to this property, as we have explicitly demonstrated, it is possible to obtain a lower dimensional sub-manifold, which is asymptotically flat. This has enabled us to apply the Geroch-Hansen formalism to derive the multipole moments of the \KN spacetime. In addition, the fact that the \KN spacetime is vacuum has also aided us in the quest to find multipole moments of this spacetime, by providing the twist potential.  

Having derived the twist potential as well as asserting the asymptotic flatness of the lower dimensional sub-manifold, the derivation of the multipole moment follows from the tensorial recursion relation derived in \cite{Geroch:1970cc,Geroch:1970cd,Hansen:1974zz}, also see \cite{Backdahl:2005be}. However, in the presence of stationarity and axi-symmetry, the above tensorial recursion relation can be reduced to a scalar recursion relation and hence it follows that the multipole moments can be derived by taking recurrent derivatives of an appropriate scalar function. Following this analysis, we have explicitly derived the multipole moments of the \KN spacetime and it has been presented in \ref{mult_mom_KerrNUT}, depicting our final result. It turns out in the limit of vanishing NUT charge, the multipole moments reduce to that of Kerr spacetime. However, in striking contrast to Kerr spacetime, for \KN spacetime both mass and spin multipole moment of all orders exist. For example, the zeroth order spin multipole moment is given by the NUT charge $N$, while the first order mass multipole moment is given by $Na$. While, for vanishing NUT charge all of these terms identically vanishes. It is quite remarkable to witness how the addition of a non-zero NUT charge changes the multipolar structure of spacetime, though the expression for the multipole moments remain compact and is consistent with a intuitive picture. As we have explored further, it turns out that non-vanishing of all the mass and spin multipole moments for \KN spacetime is a feature and not a bug, which is intimately connected with the asymmetry of the \KN spacetime about the equatorial plane. Besides, the expression for multipole moments is also consistent with the duality symmetry of the \KN spacetime, namely the multipole moments with $N=0$ is identical to the multipole moments with $M=0$, modulo a negative sign, under the following transformation, $M\leftrightarrow iN$. This provides an overview of the theoretical computations associated with the multipole moments of the \KN spacetime, which we have applied to the context of gravitational wave observations.

The observational aspects of this work stems from the analysis of the multipolar structure of the \KN spacetime. In particular, we have studied the inspiral phase of a binary system with incomparable masses, using which one can relate the multipole moments of the central massive object with the gravitational wave observables like, the radiated energy and the precession frequencies. Due to the non-vanishing contributions from the odd mass and even spin moments, it is likely that the gravitational wave observable would also contain their imprints. This is what we have achieved in this work, i.e., we have expressed the energy radiated by the gravitational waves as well as precession frequencies of the inspiraling object about the circular orbit on the equatorial plane in terms of all the multipole moments, including odd mass moments and even spin moments. Interestingly, the zeroth order spin moment appears at a leading order PN coefficient than the first order spin moment, as expected. For the \KN spacetime this translates into the fact that the NUT charge appears at a lower PN order than the rotation parameter and hence have significant influence on these observables. 

However, the above result is based on the existence of quasi-circular orbits on the equatorial plane, which are not possible on the \KN spacetime. In order to circumvent this issue, we have attempted to estimate how the value of NUT charge affects the departure of the circular orbit from the equatorial plane. As we have noticed in \ref{Figure_01}, with an increase in the value of the NUT charge, the angular deviation from the equatorial plane also increases, rendering the above analysis inappropriate. Our analysis suggests that the NUT charge should always satisfy the following bound $(N/M)<\mathcal{O}(10^{-2})$ in order to claim that the orbits are confined mostly on the equatorial plane and hence the above analysis can be carried out. In addition, we observe that circular orbits with larger radius has a smaller deviation from the equatorial plane compared to a circular orbit with smaller radius. This is expected, since an orbit with a larger radius describes an early phase of the inspiral and likely to have less interaction with the NUT charge, compared to a later stage of the inspiral. To point out another key aspect of our findings, we notice that even though the NUT charge appears at a lower PN order than the rotation parameter, due to the smallness of the ratio $(N/M)$, for most of the astrophysical situations the contribution from the NUT charge will be comparable or smaller compared to the rotation parameter. We have also provided an algorithm to read of the multipole moments from the gravitational wave observables. It would be interesting to derive the PN expansion of the observables to higher orders using the above algorithm. This would enable us to determine the effect of asymmetry about the equatorial plane from the gravitational wave observables and hence can provide further constraints on the NUT charge. This will also facilitate further constraint on the NUT charge using the inspiral part of the data from the binary black hole merger events in LIGO. These we leave for the future.   

\section*{Acknowledgement}

Both the authors thank Sukanta Bose and Naresh Dadhich for valuable discussions and for providing useful comments related to the manuscript. They are also thankful to Sayak Datta for useful conversations regarding the present topic. Finally, SM acknowledges the financial support from the Department of Science and Technology (DST), Government of India and the work of SC is supported in part by the INSPIRE Faculty fellowship (Reg. No. DST/INSPIRE/04/2018/000893) from Government of India.
\appendix
\labelformat{section}{Appendix #1} 
\labelformat{subsection}{Appendix #1}
\numberwithin{equation}{section}
\section{From spherical to cylindrical coordinate system}\label{AppSphCy}

The Kerr-NUT spacetime in the spherical coordinate system $(t,r,\theta,\phi)$ is well known. In this appendix we will demonstrate how the Kerr-NUT metric in the cylindrical coordinate system can be determined. The key to this transformation is \ref{KN_Cylindrical}. The idea is to find the coordinates $(\rho,z)$, such that the \KN metric in $(t,r,\theta,\phi)$ coordinate system expressed in \ref{metric_KN_old} can be written as \ref{KN_Cylindrical}. Equating $g_{tt}$ and $g_{t\phi}$ components of \ref{metric_KN_old} and \ref{KN_Cylindrical}, we obtain,
\begin{equation}
F=\frac{\Delta-a^{2}\sin^{2}\theta}{\Sigma^{2}}~;\quad \bar{\omega}=\frac{\Delta P-a\sin^{2}\theta (r^{2}+a^{2}+N^{2})}{\Delta -a^{2}\sin^{2}\theta}~.
\end{equation}
Therefore, by equating the $g_{\phi \phi}$ component in \ref{metric_KN_old} with that in \ref{KN_Cylindrical}, we obtain the following relation,
\begin{equation}
\frac{(r^{2}+a^{2}+N^{2})^{2}\sin^{2}\theta-\Delta P^{2}}{\Sigma^{2}}=\frac{\rho^{2}\Sigma^{2}}{\Delta-a^{2}\sin^{2}\theta}
-\frac{\Delta-a^{2}\sin^{2}\theta}{\Sigma^{2}}\left(\frac{\Delta P-a\sin^{2}\theta (r^{2}+a^{2}+N^{2})}{\Delta -a^{2}\sin^{2}\theta}\right)^{2}.
\end{equation}
The above equation can be rewritten and the cylindrical coordinate $\rho$ can be related to the spherical coordinates $(r,\theta)$ as,
\begin{align}
\rho^{2}&=\Sigma^{-4}\left\{\left(\Delta-a^{2}\sin^{2}\theta\right)\left[(r^{2}+a^{2}+N^{2})^{2}\sin^{2}\theta-\Delta P^{2} \right]+\left[\Delta P-a\sin^{2}\theta (r^{2}+a^{2}+N^{2})\right]^{2}\right\},
\nonumber
\\
&=\Delta \sin^{2}\theta~.
\end{align}
Let us now work out the connection between the other cylindrical coordinate $z$ and the spherical coordinates $(r,\theta)$. For this purpose, we assume the following decomposition, $z=f(r)\cos \theta$ and hence express the following combination in the spherical coordinates,
\begin{align}
d\rho^{2}+dz^{2}&=\left(\sqrt{\Delta}\cos \theta d\theta+\sin \theta \frac{\Delta'}{2\sqrt{\Delta}}dr\right)^{2}+\left(-f\sin \theta d\theta+f'\cos \theta dr\right)^{2},
\nonumber
\\
&=\left(\frac{\Delta'^{2}}{4\Delta}\sin^{2}\theta+f'^{2}\cos^{2}\theta\right)dr^{2}+\left(\Delta \cos^{2}\theta+f^{2}\sin^{2}\theta \right)d\theta^{2}
+2\sin \theta \cos\theta\left(\frac{\Delta'}{2}-ff'\right)drd\theta~.
\end{align}
Comparison with \ref{metric_KN_old} suggests that coefficient of the $drd\theta$ term must vanish, which yields, $\Delta'=2ff'$, which can be integrated, yielding $f(r)=\sqrt{r^{2}-2Mr+C}$, where $C$ is a constant of integration. For consistency of the $g_{rr}$ and $g_{\theta \theta}$ components with the metric presented in \ref{metric_KN_old}, we obtain, $C=M^{2}$ and hence, $z=(r-M)\cos \theta$. These are the two transformation relations used in \ref{KN_transform}. 
\section{Existence of the twist potential for vacuum spacetimes}\label{AppTwist}

In this appendix, we will demonstrate that in vacuum spacetimes there always exist a twist potential $\omega$, arising out of the twist vector field $\omega_{\mu}$ defined in \ref{KN_twistvec}. As emphasized before, the existence of the twist potential $\omega$, demands the following form for the twist vector field $\omega_{\mu}$, namely $\omega_{\mu}=\nabla_{\mu}\omega$. To see that this is really the case, we need to establish that, $\nabla_{[\mu}\omega_{\nu]}=0$. To prove the same, we start with the following relation,
\begin{equation}\label{eqapptwist1}
\nabla_{[\mu}\omega_{\nu]}=\dfrac{1}{2}\left(\nabla_{\mu}\omega_{\nu}-\nabla_{\nu}\omega_{\mu}\right)
=\dfrac{1}{2}\left(\delta^{\alpha}_{\mu}\delta^{\beta}_{\nu}-\delta^{\alpha}_{\nu}\delta^{\beta}_{\mu}\right)\nabla_{\alpha}\omega_{\beta}
=\dfrac{1}{4}\epsilon_{\mu \nu \rho \sigma}\epsilon^{\alpha \beta \rho \sigma}\nabla_{\alpha}\omega_{\beta}~,
\end{equation}
where in the last line we have used the properties of the Levi-Civita symbol $\epsilon_{\mu \nu \alpha \beta}$ appropriately. Following the definition of the twist vector field $\omega_{\mu}$, as in \ref{KN_twistvec}, we can simplify the term $\epsilon^{\rho \sigma \alpha \beta}\nabla_{\alpha}\omega_{\beta}$ as follows,
\begin{equation}
\epsilon^{\rho \sigma \alpha \beta}\nabla_{\alpha}\omega_{\beta}
=\epsilon^{\rho \sigma \alpha \beta}\nabla_{\alpha}\left(\epsilon_{\beta \eta \sigma \delta}\xi^{\eta}_{(t)}\nabla^{\sigma}\xi^{\delta}_{(t)}\right)
=\epsilon^{\rho \sigma \alpha \beta}\epsilon_{\beta \eta \sigma \delta}\nabla_{\alpha}\left(\xi^{\eta}_{(t)}\nabla^{\sigma}\xi^{\delta}_{(t)}\right)~.
\end{equation}
By using the properties of the Levi-Civita tensor and the fact that $\xi^{\alpha}_{(t)}$ is a Killing vector field, we can express the above relation as,  
\begin{equation}\label{eq:appendix_01}
\epsilon^{\rho \sigma \alpha \beta}\nabla_{\alpha}\omega_{\beta}=-2\left\{\nabla_{\alpha}\left(\xi^{\alpha}_{(t)}\nabla^{\rho}\xi^{\sigma}_{(t)}\right)
+\nabla_{\alpha}\left(\xi^{\sigma}_{(t)}\nabla^{\alpha}\xi^{\rho}_{(t)}\right)+\nabla_{\alpha}\left(\xi^{\rho}_{(t)}\nabla^{\sigma}\xi^{\alpha}_{(t)}\right)\right\}~.
\end{equation}
The above expression can be further simplified using the properties of the Killing vector field $\xi^{\alpha}_{(t)}$, e.g., we have the following identity,
\begin{eqnarray}
\nabla_{\alpha}\left(\xi^{\alpha}_{(t)}\nabla^{\rho}\xi^{\sigma}_{(t)}\right)
&=&\left(\nabla_{\alpha}\xi^{\alpha}_{(t)}\right)\left(\nabla^{\rho}\xi^{\sigma}_{(t)}\right)+\xi^{\alpha}_{(t)}\left(\nabla_{\alpha}\nabla^{\rho}\xi^{\sigma}_{(t)}\right)=0~,
\end{eqnarray}
where, we have used the result $\nabla_{\alpha}\xi^{\alpha}_{(t)}=0$ and the following identity,
\begin{equation}
\nabla^{\alpha}\nabla_{\beta}\xi_{\rho}^{(t)}=-R^{\alpha}_{~\sigma \beta \rho}\xi^{\sigma}_{(t)}~, 
\label{eq:appendix_02}
\end{equation}
as well as the antisymmetry of the Riemann tensor in the first two indices. Therefore, \ref{eq:appendix_01} simplifies to the following form,
\begin{equation}
\epsilon^{\rho \sigma \alpha \beta}\nabla_{\alpha}\omega_{\beta}=-2\left\{\nabla_{\alpha}\left(\xi^{\sigma}_{(t)}\nabla^{\alpha}\xi^{\rho}_{(t)}\right)+\nabla_{\alpha}\left(\xi^{\rho}_{(t)}\nabla^{\sigma}\xi^{\alpha}_{(t)}\right)\right\}~.
\end{equation}
Expanding out the derivatives and using the relation, $\nabla_{\mu}\xi_{\nu}^{(t)}+\nabla_{\nu}\xi_{\mu}^{(t)}=0$, we obtain,
\begin{align}
\epsilon^{\rho \sigma \alpha \beta}\nabla_{\alpha}\omega_{\beta}&=-2\left\{\xi^{\sigma}_{(t)}\nabla_{\alpha}\left(\nabla^{\alpha}\xi^{\rho}_{(t)}\right)
+\xi^{\rho}_{(t)}\nabla_{\alpha}\left(\nabla^{\sigma}\xi^{\alpha}_{(t)}\right)\right\},
\nonumber
\\
&=2\left\{\xi^{\sigma}_{(t)}R^{\rho}_{\alpha}\xi^{\alpha}_{(t)}-\xi^{\rho}_{(t)}R^{\sigma}_{\alpha}\xi^{\alpha}_{(t)}\right\}~,
\end{align}
where, in arriving at the last line we have used \ref{eq:appendix_02}. Therefore, \ref{eqapptwist1}, simplifies to,
\begin{equation}
\nabla_{[\mu}\omega_{\nu]}=\dfrac{1}{2}\epsilon_{\mu \nu \rho \sigma}\left\{\xi^{\sigma}_{(t)}R^{\rho}_{\alpha}\xi^{\alpha}_{(t)}-\xi^{\rho}_{(t)}R^{\sigma}_{\alpha}\xi^{\alpha}_{(t)}\right\}
=\epsilon_{\mu \nu \rho \sigma}\xi^{\sigma}_{(t)}R^{\rho}_{\alpha}\xi^{\alpha}_{(t)}.
\end{equation}
Therefore, for the vacuum solutions and for maximally symmetric spacetimes, the above expression identically vanishes, thereby ensuring existence of the twist potential $\omega$. 

\section{Connection with Thorne's Formalism}\label{AppThorne}

In this section, we expand the $g_{00}$ and $g_{0\phi}$ components of the \KN metric in an asymptotic series, from which the initial multipole moments can be determined. Following \cite{Thorne:1980ru}, we introduce the tetrads, $e^{\mu}_{0}=\partial_{t}$ and $e^{\mu}_{\phi}=(r\sin \theta)^{-1}\partial_{\phi}$ and hence we have the following asymptotic expansion for the metric coefficients, 
\begin{align}
g_{00}&=e_{0}^{\mu}e_{0}^{\nu}g_{\mu \nu}=-\frac{\Delta-a^{2}\sin^{2}\theta}{\Sigma^{2}}=\frac{-r^{2}+2Mr-a^{2}\cos^{2}\theta+N^{2}}{r^{2}+\left(N+a\cos\theta\right)^{2}}
\nonumber
\\
&=\left(-1+\frac{2M}{r}-\frac{a^{2}\cos^{2}\theta}{r^{2}}+\frac{N^{2}}{r^{2}}\right)\left(1+\frac{\left(N+a\cos\theta\right)^{2}}{r^{2}}\right)^{-1}
\nonumber
\\
&=\left(-1+\frac{2M}{r}-\frac{a^{2}\cos^{2}\theta}{r^{2}}+\frac{N^{2}}{r^{2}}\right)\left(1-\frac{\left(N+a\cos\theta\right)^{2}}{r^{2}}+\frac{\left(N+a\cos\theta\right)^{4}}{r^{4}}\right)
\nonumber
\\
&=-1+\frac{2M}{r}+\frac{2Na\cos\theta+2N^{2}}{r^{2}}-\frac{2M\left(N+a\cos\theta\right)^{2}}{r^{3}}-\frac{a^{2}N^{2}\cos^{2}\theta}{r^{4}}+\mathcal{O}\left(\frac{1}{r^{5}}\right)~,
\end{align}
as well as
\begin{align}
g_{0\phi}&=\frac{g_{t\phi}}{r\sin \theta}=\frac{\Delta P-a\sin^{2}\theta (r^{2}+a^{2}+N^{2})}{r\sin \theta\Sigma^{2}}
\nonumber
\\
&=\frac{\left(r^{2}-2Mr+a^{2}-N^{2}\right)\left(a\sin^{2}\theta-2N\cos\theta\right)-a\sin^{2}\theta (r^{2}+a^{2}+N^{2})}{r\sin \theta\Sigma^{2}}
\nonumber
\\
&=\frac{\left(-2Mr-2N^{2}\right)a\sin^{2}\theta-2N\left(r^{2}-2Mr+a^{2}-N^{2}\right)\cos\theta}{r^{3}\sin \theta}\left(1+\frac{\left(N+a\cos\theta\right)^{2}}{r^{2}}\right)^{-1}
\nonumber
\\
&=\frac{1}{r^{2}}\left[\left(-2M-\frac{2N^{2}}{r}\right)a\sin\theta-2N\left(r-2M+\frac{a^{2}-N^{2}}{r}\right)\cot\theta \right]\left(1-\frac{\left(N+a\cos\theta\right)^{2}}{r^{2}}+\frac{\left(N+a\cos\theta\right)^{4}}{r^{4}}\right)
\nonumber
\\
&=-\frac{2N\cot\theta}{r}-\frac{2Ma\sin\theta+4NM\cot\theta}{r^{2}}+\mathcal{O}\left(\frac{1}{r^{3}}\right).
\end{align}
These expressions have been referred to in the main text, while relating the multipole moments from the Geroch-Hansen formalism with the Thorne's formalism. 

\section{Expressions for the circular orbits on the equatorial plane}\label{AppCirc}

In order to derive the following results, we have employed the notations and expressions given in \cite{Mukherjee:2018dmm}. From Eq. [12] of \cite{Mukherjee:2018dmm}, the expression of $\dot{\theta}$ is given as follows: 
\begin{eqnarray}
\dot{\theta}=\dfrac{\Bigl\{\lambda^2 \sin^2\theta+(L_{\rm z}-a E)^2-(aE \sin^2\theta-2 l \cos\theta)-L_{\rm z})^2-\sin^2\theta (l+a \cos\theta)^2))\Bigr \}^{1/2}}{\sin\theta \left [r^2_{\rm c}+(l+a \cos\theta)^2 \right]},
\end{eqnarray}
where, $r_{\rm c}$, $\lambda$, $E$, $L_{\rm z}$ has the usual meaning of radius of the orbit, Carter constant, energy, angular momentum associated with the orbit. By differentiating with the proper time, we obtain $\ddot{\theta}$ and is given by
\begin{eqnarray}
\ddot{\theta}&=&\dfrac{1}{\left[r_{\rm c}^2+(l+a\cos\theta)^2\right]^3}\Bigl\{-4 r_{\rm c} \Bigl(-(L_{\rm z}-aE)^2(r^2_{\rm c}-2 M r_{\rm c}+a^2-l^2)+((r^2_{\rm c}+a^2+l^2)E-a L_{\rm z})^2 \nonumber \\
& & -(r^2_{\rm c}-2 M r_{\rm c}+a^2-l^2)(r^2_{\rm c}+\lambda) \Bigr)+(r^2_{\rm c}+(l+a \cos\theta)^2)\Bigl[2(L_{\rm z}-aE)^2(M-r_{\rm c})-\nonumber \\
& & 2 r_{\rm c} (r^2_{\rm c}-2 Mr_{\rm c}+a^2-l^2)+4E r_{\rm c} (E(r^2_{\rm c}+a^2+l^2)-aL_{\rm z})+2(M-r_{\rm c})(r^2_{\rm c}+\lambda)\Bigr]\Bigr\}.
\end{eqnarray}
From the condition of timelike circular orbit, i.e., $\dot{r}=\ddot{r}=0$, the expression for energy is given as ( Eq. [39] in \cite{Mukherjee:2018dmm})
\begin{equation}
{E}_{c}^{+}=(1+l^2 u^2_{\rm c})^{-1}\Big[1+l^2 u^2_{\rm c}-2 M u_{\rm c} (1+l^2 u^2_{\rm c})-a^2 u^2_{\rm c}(1+\lambda u^2_{\rm c})+   a u^{3/2}_{\rm c}\mathcal{K}^{1/2}\Big] \left\{Z_{+}^{c}(1+l^2u_{c}^2)\right\}^{-1/2},
\label{energy_circular}
\end{equation}
where, we assume the orbit to be co-rotating and $u_{\rm c}=1/r_{\rm c}$. The expression for angular momentum is given as
\begin{equation}
L_{\rm z}=aE+x_{+}^{c}.
\end{equation}
The expression for $x_{+}^{c}$ is given by
\begin{equation}
x_{+}^{c}=-\dfrac{1}{{u_{\rm c} Z_{+}^{c}}^{1/2}}\dfrac{1+\lambda u^2_{\rm c}}{(1+l^2 u_{\rm c}^2)^{1/2}}\left\{a\sqrt{u_{\rm c}}- (1+\lambda u^2_{\rm c})^{-1}\mathcal{K}^{1/2}\right\},
\end{equation}
where $Z_{+}^{c}$ and $\mathcal{K}$, has the following expressions
\begin{equation}
Z_{+}^{c}=(1+l^2 u_{\rm c}^2)^{-1}\Big[\left\{1-l^4 u_{\rm c}^4 +M u_{\rm c}(l^4 u^4_{\rm c}-2 l^2 u^2_{\rm c}-3)-2 a^2 u^2_{\rm c} (1+\lambda u^2_{\rm c})\right\}+ 2au^{3/2}_{\rm c}\mathcal{K}^{1/2}\Big],
\end{equation} 
and, 
\begin{equation}
\mathcal{K}=M(1+l^2 u^2_{\rm c})\left\{1+3 \lambda u^2_{\rm c}-3 l^2 u^2_{\rm c}-\lambda l^2 u^4_{\rm c}\right\}+u_{\rm c} \left\{2 l^2-\lambda+l^4 u^2_{\rm c}(2+u^2_{\rm c} \lambda)+a^2 (1+\lambda u^2_{\rm c})^2\right\}.
\end{equation}

\section{Deriving the Recursion Relation}\label{AppRec}

In this appendix we will derive the recursion relation relevant for obtaining the multipole moments from the gravitational wave observations. In this context the Ernst potential $\varepsilon$, defined in \ref{ernst_pot} plays an important role, as it satisfies the following differential equation (see \cite{stephani2009exact}), 
\begin{equation}\label{eq:appen_01}
\boldsymbol{\nabla}^2\varepsilon = \dfrac{1}{\lambda}(\boldsymbol{\nabla}\varepsilon)\cdot (\boldsymbol{\nabla}\varepsilon)~,
\end{equation}
where the differential operator $\boldsymbol{\nabla}$ is the three dimensional directional derivative on the manifold $M_{3}$. Arising out of the Ernst potential is the quantity $\Phi$, defined in \ref{mult_pot}, which plays a central role in the analysis of the multipole moments. The differential equation satisfied by the potential $\Phi$ can be determined along the following lines, first of all,
\begin{equation}
\boldsymbol{\nabla}^{2} \Phi=\boldsymbol{\nabla}^2\left(\dfrac{1-\varepsilon}{1+\varepsilon}\right)
=-2\dfrac{\boldsymbol{\nabla}^2 \varepsilon}{(1+\varepsilon)^2}+\dfrac{4\boldsymbol{\nabla}\varepsilon \cdot \boldsymbol{\nabla}\varepsilon}{(1+\varepsilon)^3}~.
\end{equation}
By using \ref{eq:appen_01}, the above equation can be rewritten as follows,
\begin{eqnarray}
\boldsymbol{\nabla}^{2}\Phi&=&-\dfrac{2}{\lambda}\dfrac{\boldsymbol{\nabla}\varepsilon \cdot \boldsymbol{\nabla}\varepsilon}{(1+\varepsilon)^{2}}
+\dfrac{4\boldsymbol{\nabla}\varepsilon \cdot \boldsymbol{\nabla}\varepsilon}{(1+\varepsilon)^3}
\nonumber 
\\
&=&-\dfrac{2}{\lambda}\dfrac{\boldsymbol{\nabla}\varepsilon \cdot \boldsymbol{\nabla}\varepsilon}{(1+\varepsilon)^3} \left(1+\varepsilon-2 \lambda\right) 
\nonumber 
\\
& = & -\dfrac{(1+\lambda+i \omega)(1-\lambda+i \omega)}{2\lambda}\left(\boldsymbol{\nabla}\Phi \cdot \boldsymbol{\nabla}\Phi \right)~.
\label{eq:appen_02}
\end{eqnarray}
In arriving at the last line, we have used the following result, 
\begin{align}
\boldsymbol{\nabla}\Phi \cdot \boldsymbol{\nabla}\Phi&=\left(\frac{-\boldsymbol{\nabla}\varepsilon}{1+\varepsilon}-\frac{1-\varepsilon}{(1+\varepsilon)^{2}}\boldsymbol{\nabla}\varepsilon \right)\cdot \left(\frac{-\boldsymbol{\nabla}\varepsilon}{1+\varepsilon}-\frac{1-\varepsilon}{(1+\varepsilon)^{2}}\boldsymbol{\nabla}\varepsilon \right)
\nonumber
\\
&=\frac{4}{(1+\varepsilon)^{4}}\left(\boldsymbol{\nabla}\varepsilon \cdot \boldsymbol{\nabla}\varepsilon\right)~.
\end{align}
Besides, we can also have the following result related to the $\Phi$ and its complex conjugate $\Phi^{*}$ as follows,
\begin{equation}
(\Phi \Phi^{*})^{-1}=\dfrac{(1+\lambda+i \omega)(1+\lambda-i \omega)}{(1-\lambda-i \omega)(1-\lambda+i \omega)}=\dfrac{(1+\lambda)^2+\omega^2}{(1-\lambda)^2+\omega^2}~,
\end{equation}
which upon further simplification yields,
\begin{equation}
\Phi \Phi^{*}-1=\dfrac{-4\lambda}{(1+\lambda)^2+\omega^2}~.
\end{equation}
Thus we immediately obtain,
\begin{equation}
\dfrac{2\Phi^{*}}{|\Phi|^{2}-1}=\dfrac{2(1-\lambda+i \omega)(1+\lambda+i \omega)}{-4\lambda}~.
\end{equation}
Therefore, \ref{eq:appen_02} becomes
\begin{equation}
\boldsymbol{\nabla}^{2}\Phi=\dfrac{2\Phi^{*}}{|\Phi|^2-1}\boldsymbol{\nabla}\Phi \cdot \boldsymbol{\nabla}\Phi~,
\end{equation}
which is the desired differential equation for $\Phi$. However, it is often advantageous to introduce a set of new coordinates $\tilde{\rho}$ and $\tilde{z}$ from the old cylindrical coordinates $(\rho,z)$ introduced in \ref{KN_Cylindrical}, such that, 
\begin{equation}
\tilde{\rho}=\frac{\rho}{\rho^{2}+z^{2}}~;\quad \tilde{z}=\frac{z}{\rho^{2}+z^{2}}~;\quad \tilde{r}=\sqrt{\tilde{\rho}^{2}+\tilde{z}^{2}}~.
\end{equation}
Asymptotically, $\tilde{\rho}$ and $\tilde{z}$ coincides with the $\bar{\rho}$ and $\bar{z}$ defined in \ref{bar_coord}. Thus the unphysical potential becomes, $\bar{\Phi}=\Phi/\sqrt{\Omega}=(1/\tilde{r})\Phi$ and it satisfies the following differential equation,
\begin{equation}\label{diffeq_unphys}
\left(\tilde{r}^{2}\bar{\Phi}\bar{\Phi}^{*}-1\right)\boldsymbol{\nabla}^{2}\bar{\Phi}=2\bar{\Phi}^{*}\left[\tilde{r}^{2}\left(\boldsymbol{\nabla}\bar{\Phi}\right)^{2}
+2\tilde{r}\bar{\Phi}\boldsymbol{\nabla}\bar{\Phi}\cdot \boldsymbol{\nabla}\tilde{r}+\left(\boldsymbol{\nabla}\tilde{r}\right)^{2}\bar{\Phi}^{2}\right]~.
\end{equation}
From \ref{expansion}, the expansion of the unphysical potential $\bar{\Phi}$ in the new $(\tilde{\rho},\tilde{z})$ coordinate system is given by,
\begin{equation}\label{app_expansion}
\bar{\Phi}=\sum_{\substack{i=0 \\ j = 0}}a_{ij}\tilde{\rho}^{i}\tilde{z}^{j}~.
\end{equation}
This expansion must be substituted in \ref{diffeq_unphys}, in order to determine the coefficients $a_{ij}$. In the coordinate system $(\tilde{\rho},\tilde{z},\phi)$ we obtain the following expression for the Laplacian operator on the three dimensional manifold $M_{3}$, 
\begin{equation}
\boldsymbol{\nabla}^{2}\bar{\Phi}=e^{-2\gamma}\left[\partial^{2}_{\tilde{\rho}}\bar{\Phi}+\dfrac{1}{\tilde{\rho}}\partial_{\tilde{\rho}}\bar{\Phi}+\partial^{2}_{\tilde{z}}\bar{\Phi}\right],
\end{equation}
where, $\partial^{2}_{\phi}\bar{\Phi}$ term is absent as $\bar{\Phi}$ is independent of the coordinate $\phi$, thanks to the axi-symmetry of the compact object. Substituting \ref{app_expansion} in the above expression, we obtain,
\begin{equation}
\boldsymbol{\nabla}^{2}\bar{\Phi}=e^{-2\gamma} \left[\sum_{\substack{i=2 \\ j = 0}}i(i-1)a_{ij}\tilde{\rho}^{i-2}\tilde{z}^{j}
+\sum_{\substack{i=1 \\ j = 0}}ia_{ij}\tilde{\rho}^{i-2}\tilde{z}^{j}+\sum_{\substack{i=0 \\ j = 2}}j(j-1)a_{ij}\tilde{\rho}^{i}\tilde{z}^{j-2}\right].
\label{eq:nabla_appen}
\end{equation}
In addition, we also have the following relations in the three dimensional manifold $M_{3}$, which take the following forms,
\begin{eqnarray}
\boldsymbol{\nabla}\bar{\Phi} \cdot \boldsymbol{\nabla}\bar{\Phi}&=&e^{-2\gamma}\left[(\partial_{\tilde{\rho}}\bar{\Phi})^{2}
+(\partial_{\tilde{z}}\bar{\Phi})^{2}\right]
\nonumber 
\\
\boldsymbol{\nabla}\bar{\Phi}\cdot \boldsymbol{\nabla}\tilde{r}&=&e^{-2\gamma}\left[\dfrac{\tilde{\rho}}{\tilde{r}}\partial_{\tilde{\rho}}\bar{\Phi}
+\dfrac{\tilde{z}}{\tilde{r}}\partial_{\tilde{z}}\bar{\Phi}\right] 
\nonumber 
\\
\boldsymbol{\nabla}\tilde{r}\cdot \boldsymbol{\nabla}\tilde{r}&=&e^{-2\gamma}~.
\end{eqnarray}
Therefore, we can write \ref{diffeq_unphys} as follows,
\begin{equation}
\left[(\tilde{\rho}^{2}+\tilde{z}^{2})\bar{\Phi}\bar{\Phi}^{*}-1\right]\left[\partial^{2}_{\tilde{\rho}}\bar{\Phi}+\dfrac{1}{\tilde{\rho}}\partial_{\tilde{\rho}}\bar{\Phi}+\partial^{2}_{\tilde{z}}\bar{\Phi}\right]=2\bar{\Phi}^{*}\left[(\tilde{\rho}^2+\tilde{z}^2)\left\{(\partial_{\tilde{\rho}}\bar{\Phi})^{2}+(\partial_{\tilde{z}}\bar{\Phi})^{2}\right\}+2 \bar{\Phi}\left\{\tilde{\rho}\partial_{\tilde{\rho}}\bar{\Phi}+\tilde{z}\partial_{\tilde{z}}\bar{\Phi}\right\}+\bar{\Phi}^{2}\right]~,
\end{equation}
which can be rewritten as,
\begin{align}
\partial^{2}_{\tilde{\rho}}\bar{\Phi}+\dfrac{1}{\tilde{\rho}}\partial_{\tilde{\rho}}\bar{\Phi}+\partial^{2}_{\tilde{z}}\bar{\Phi}
&=\left(\tilde{\rho}^{2}+\tilde{z}^{2}\right)\bar{\Phi}\bar{\Phi}^{*}\left(\partial^{2}_{\tilde{\rho}}\bar{\Phi}+\dfrac{1}{\tilde{\rho}}\partial_{\tilde{\rho}}\bar{\Phi}+\partial^{2}_{\tilde{z}}\bar{\Phi} \right)
\nonumber
\\
&\hskip 1 cm -2\bar{\Phi}^{*}\left[(\tilde{\rho}^2+\tilde{z}^2)\left\{(\partial_{\tilde{\rho}}\bar{\Phi})^{2}+(\partial_{\tilde{z}}\bar{\Phi})^{2}\right\}+2 \bar{\Phi}\left\{\tilde{\rho}\partial_{\tilde{\rho}}\bar{\Phi}+\tilde{z}\partial_{\tilde{z}}\bar{\Phi}\right\}+\bar{\Phi}^{2}\right]~.
\end{align}
Using the expansion of $\bar{\Phi}$ and \ref{eq:nabla_appen}, the above differential equation becomes, 
\begin{align}
\sum_{p,q}a_{pq}p(p-1)\tilde{\rho}^{p-2}\tilde{z}^q&+\sum_{p,q}a_{pq}p \tilde{\rho}^{p-2}\tilde{z}^{q}+\sum_{p,q}a_{pq}q(q-1)\tilde{\rho}^{p}\tilde{z}^{q-2}  
\nonumber
\\
&\hskip -2 cm =\left(\tilde{\rho}^{2}+\tilde{z}^{2}\right)\sum a_{kl}a^{*}_{mn}\tilde{\rho}^{k+m}\tilde{z}^{l+n}
\nonumber 
\\
&\times \Big[\sum a_{pq}p(p-1)\tilde{\rho}^{p-2}\tilde{z}^q+\sum a_{pq}p\tilde{\rho}^{p-2}\tilde{z}^{q}+\sum a_{pq}q(q-2)\tilde{\rho}^{p}\tilde{z}^{q-2}\Big]
\nonumber 
\\
&\hskip -2 cm -4\sum a_{kl}a^{*}_{mn}\tilde{\rho}^{k+m}\tilde{z}^{l+n} \Big(\sum a_{pq}p\tilde{\rho}^{p}\tilde{z}^q+\sum a_{pq}q\tilde{\rho}^{p}\tilde{z}^{q}\Big)
-2 \sum a_{kl}a^{*}_{mn}a_{pq}\tilde{\rho}^{k+m+p}\tilde{z}^{l+n+q}
\nonumber
\\
&\hskip -2 cm -2 \sum a^{*}_{mn} \tilde{\rho}^{m}\tilde{z}^{n}\left(\tilde{\rho}^{2}+\tilde{z}^{2}\right)\Big[\Big(\sum a_{pq}p\tilde{\rho}^{p-1}\tilde{z}^{q}\Big)
\times \Bigl(\sum a_{kl}k\tilde{\rho}^{k-1}\tilde{z}^{l}\Bigr)
\nonumber 
\\
&\hskip 3 cm +\Bigl(\sum a_{pq}q\tilde{\rho}^{p}\tilde{z}^{q-1}\Bigr)\Bigl(\sum a_{kl}l\tilde{p}^k\tilde{z}^{l-1}\Bigr)\Bigr\}~.
\end{align}
The above equation can be simplified further by renaming the indices appropriately, which yields,
\begin{align}
\sum a_{p+2,q} \tilde{\rho}^{p}\tilde{z}^{q}(p+2)^{2}&+\sum a_{p,q+2}(q+2)(q+1)\tilde{\rho}^{p}\tilde{z}^{q}
\nonumber
\\
&\hskip -2 cm =\sum a_{kl}a^{*}_{mn}\tilde{\rho}^{k+m}\tilde{z}^{l+n}\Big[\sum a_{pq}p^2 \Bigl(\tilde{\rho}^p \tilde{z}^q+\tilde{\rho}^{p-2}\tilde{z}^{q+2}\Bigr)
+\sum a_{pq} q(q-1)\Bigl(\tilde{\rho}^{p+2}\tilde{z}^{q-2}+\tilde{\rho}^p \tilde{z}^q \Bigr)\Big]
\nonumber \\
&\hskip -2 cm -4 \sum a_{kl}a^{*}_{mn}\tilde{\rho}^{k+m}\tilde{z}^{l+n}\Big(\sum p a_{pq} \tilde{\rho}^p \tilde{z}^q+\sum q a_{pq}\tilde{\rho}^p \tilde{z}^q\Big)
-2 \sum a_{kl}a^{*}_{mn}a_{pq}\tilde{\rho}^{k+m+p}\tilde{z}^{l+n+q}
\nonumber \\
&\hskip -2 cm -2 \sum a^{*}_{mn}\tilde{\rho}^{m}\tilde{z}^{n}(\tilde{\rho}^2+\tilde{z}^2) \left(\sum pk a_{pq}a_{kl}\tilde{\rho}^{p+k-2}\tilde{z}^{l+q}
+\sum lq~a_{pq}a_{kl}\tilde{\rho}^{p+k}\tilde{z}^{l+q-2}\right)
\end{align}
Relabelling the indices and then equating the coefficients of equal powers of $\tilde{\rho}$ and $\tilde{z}$, we obtain the following recursion relation for determining the coefficients $a_{jk}$,
\begin{align}
(r+2)^{2}a_{r+2,s}&=-(s+1)(s+2)a_{r,s+2}
\nonumber
\\
&+\sum_{\substack{k+m+p=r \\ l+n+q=s}}a_{kl}a^{*}_{mn}\Bigl[a_{p+2,q-2}(p+2)(p+2-2k)+a_{p-2,q+2}(q+2)(q+1-2l)
\nonumber 
\\
&\hskip 4 cm +a_{pq}(p^2+q^2-4p-5q-2pk-2ql-2)\Bigr]~,
\end{align}
which have been used in the main text. 

\bibliography{References}
\bibliographystyle{./utphys1}
\end{document}